\documentclass[showpacs,preprintnumbers,amsfonts,pre]{revtex4}
\usepackage{graphicx}% Include figure files
\usepackage{bm}% bold math
\begin{document}
\title{
Punctuated equilibria and $1/f$ noise in a  
biological coevolution model with individual-based dynamics
}
\author{Per Arne Rikvold$^{1,2,3,}$}\email{rikvold@csit.fsu.edu}
%and 
\author{R.K.P.\ Zia$^{2,}$}\email{rkpzia@vt.edu}
\affiliation{
$^1$School of Computational Science and Information Technology, 
Center for Materials Research and Technology, and Department of Physics,
Florida State University, Tallahassee, Florida 32306-4120\\
$^2$Center for Stochastic Processes in Science and Engineering, 
Department of Physics, Virginia Polytechnic Institute and State University,
Blacksburg, Virginia 24061-0435\\
$^3$Department of Physics and Astronomy and ERC Center for Computational
Sciences, Mississippi State University, Mississippi State, 
Mississippi 39762-5167
}
\date{\today}

\begin{abstract}
We present a study by linear stability analysis and large-scale Monte
Carlo simulations of a simple model of biological coevolution. Selection 
is provided through a reproduction probability that contains 
quenched, random interspecies interactions, while genetic variation is 
provided through a low mutation rate. Both selection and mutation 
act on individual organisms. 
Consistent with some current theories of macroevolutionary dynamics, 
the model displays intermittent, statistically self-similar
behavior with punctuated equilibria. 
The probability density for the lifetimes of ecological 
communities is well approximated by a power law with exponent
near $-2$, and the corresponding power spectral
densities show $1/f$ noise (flicker noise) over several decades. 
The long-lived communities (quasi-steady states) consist of  
a relatively small number of mutualistically interacting species, 
and they are surrounded by a ``protection zone'' of closely related genotypes 
that have a very low probability of invading the resident community. 
The extent of the protection zone affects the stability of the community
in a way analogous to the height of the free-energy barrier surrounding
a metastable state in a physical system. 
Measures of biological diversity are on average stationary 
with no discernible trends, even over our 
very long simulation runs of approximately $3.4 \times 10^7$ generations. 
\end{abstract}


\maketitle

\section{Introduction}
\label{sec:Int}

Biological evolution offers a number of important, unsolved problems that are 
well suited for investigation by methods from statistical physics. Many
of these can be studied using complex, 
interacting model systems far from equilibrium \cite{DROS01}. 
Areas that have generated exceptional interest among physicists
are those of coevolution and speciation. 
A large class of coevolution models have been inspired by one
introduced by Bak and Sneppen \cite{BAK93}, in which species with
different levels of ``fitness" compete, and the least fit species and
those that interact with it are regularly ``mutated" and replaced by new
species with different, randomly chosen fitness. Models in this class 
exhibit avalanches of extinctions and appear to evolve towards a
self-organized critical state \cite{BAK87,BAK88}. Although such models may be
said to incorporate Darwin's principle of ``survival of the fittest," 
they are artificial in the sense that mutation and selection are assumed
to act collectively on entire species, 
rather than on individual members of their populations.  

In reality, mutations are changes in the genotypes of 
{\it individual organisms\/} that are introduced or passed on 
during reproduction. These changes in the genotype affect the phenotype
(physical and behavioral characteristics of the organism and its
interactions with other organisms), and it is on the
level of the phenotypes of individuals that competition and selection act.
A number of evolution models (see, e.g., Ref.~\cite{DROS01} for a
review from a physicist's point of view) therefore take as their
basis a genome in the form of a string of ``letters" as in Eigen's 
quasi-species model of molecular evolution \cite{EIGE71,EIGE88}. 
Depending on the level of the modeling, the letters of the genomic
``alphabet'' can be a (possibly large) number of alleles at a gene
locus, or the four nucleotides of a DNA or RNA sequence \cite{BAAK99}. 
However, the size of the alphabet is not of great importance in
principle, and it is common in  
models to use a binary alphabet with the two letters 0 and 1 (or $\pm1$) 
\cite{DROS01,EIGE71,EIGE88,BAAK99}. 
%Although models have been introduced, in which a genotype in the form of a 
%string of binary digits determine a phenotype on which selection can act
%\cite{TAYL00}, for reasons of computational efficiency it is common in models 
%of macroevolution to dispense with the phenotype level and let selection act
%directly on the genotype, which is often considered to evolve in a fixed
%``fitness landscape" that may be either rugged or smooth
%\cite{DROS01,EIGE71,EIGE88,BAAK99}. 

We believe a fruitful approach to the study of coevolution is one in
which selection is provided by interspecies interactions along lines
commonly considered in community ecology, while genetic variation
is provided by random mutations in the genomes of individual organisms. 
An early attempt in this direction
is the coupled $NK$ model with population dynamics introduced by Kauffman and 
Johnsen \cite{KAUF91,KAUF93},
but thus far not many similar models have been studied.
Recently, Hall, Christensen, and collaborators \cite{HALL02,CHRI02,COLL03}
introduced a model they called the ``tangled-nature'' model, in which each
individual lives in a dynamically evolving ``fitness landscape" created
by the populations of all the other species.  Here 
we consider a simplified version of this model,
in which no individual is allowed to live through more than one 
reproduction cycle. 
This restriction to nonoverlapping generations enables us to both study 
the model in detail by linear stability analysis and to perform very long
Monte Carlo simulations of its evolutionary behavior. In short, the model
consists of populations of different species, 
on which selection acts through asexual reproduction rates that
depend on the populations of all the other species via a constant,
random interaction matrix 
%(elements chosen at random initially, but fixed for all subsequent times) 
that allows mutualistic, competitive, and
predator/prey relations. In addition to these direct interactions, all
individuals interact indirectly through competition
for a shared resource. The competition 
keeps the total population from diverging. 
Genetic variety is provided by a low mutation rate that acts on the
genomes of individual organisms during reproduction, inducing the
populations to move through genotype space. The resulting
motion is intermittent:
long periods of stasis with only minor fluctuations are interrupted by bursts
of significant change that is rapid on a macroscopic time scale,  
reminiscent of what is known in evolutionary biology as 
punctuated equilibria \cite{GOUL77,GOUL93,NEWM85}.
A preliminary report on some aspects of the work reported here is given in 
Ref.~\cite{RIKV03A}. 

The main foci of the present paper are the structure and stability of
communities and the statistical properties of the dynamical
behavior on very long (``geological'') time scales. 
The rest of the paper is organized as follows. In Sec.~\ref{sec:Mod} we
describe our model in detail, including the detailed Monte Carlo algorithm
used for our simulations. In Sec.~\ref{sec:LSA} we discuss the properties of
the fixed points of the population for the mutation-free version of the
model. Many of these properties can be understood analytically within a
simple mean-field approach. A full, probabilistic description of our model
is also possible, although the mathematics is somewhat involved. In  
Appendix~\ref{sec:AB}
we provide, for the sake of completeness, the key equations of
this approach. In Sec.~\ref{sec:MC} we give 
a detailed report on our large-scale Monte
Carlo simulations and the numerical results, together with a discussion of
their relations to the fossil record and to other theoretical models. 
Finally, in Sec.~\ref{sec:Concl}, we summarize our results and
give our conclusions and suggestions for future studies.

\section{Model and algorithm}
\label{sec:Mod}

As mentioned in Sec.~\ref{sec:Int}, 
we use a simplified version of the tangled-nature model 
introduced by Hall, Christensen, and collaborators 
\cite{HALL02,CHRI02,COLL03}. 
It consists of a population of individuals with a genome of $L$ genes,  
each of which can take one of the two values 0 or 1. Thus, the total number of 
different genotypes is ${\cal N}_{\rm max} = 2^L$. We consider each different 
genotype as a separate species, and we shall in this paper use the two terms
interchangeably. This is justified by the idea that in 
the relatively short genomes we can consider computationally, each binary 
``gene'' actually represents a group of genes in a coarse-grained sense. 

In our version of the model, 
the population evolves in discrete, nonoverlapping generations 
(as in, e.g., many insects), and 
the number of individuals of genotype $I$ in generation $t$ is $n_I(t)$. 
The total population is $N_{\rm tot}(t) = \sum_I n_I(t)$. 
In each generation, the probability that an individual of genotype $I$ produces
a litter of $F$ offspring before it dies 
is $P_I(\{n_J(t)\})$, while the probability that it 
dies without offspring is $1-P_I$. 
Although the fecundity $F$ could be quite complex in reality (e.g., a
function of $I$ on the average, but random both in time and for each
individual), we here take a simplistic approach and assume it is a
constant, independent of $I$ and $t$.
The main difference from the model of Refs.~\cite{HALL02,CHRI02,COLL03} is
that the generations are nonoverlapping in our model, while
individuals in the original model may live through several successive
reproduction cycles. This simplification facilitates theoretical analysis 
and enables us to make significantly longer simulations than those reported in
Refs.~\cite{HALL02,CHRI02,COLL03}. 
In the ``opposite'' direction, the earlier model could
be generalized to include non-trivial ``age structures,'' since
individuals living through several cycles can be assigned an age, with
age-dependent survival and reproductive properties 
\cite{LESL45,PIEL69,CHAR94,PENN95,HOWA01,ZIA02,TUZE01A,CHOW03A,CHOW03B}. 

As in the original model \cite{HALL02,CHRI02,COLL03}, 
the reproduction probability $P_I$ is taken as
\begin{equation}
P_I(\{n_J(t)\})
= 
\frac{1}{1 + \exp\left[ - \sum_J M_{IJ} n_J(t)/N_{\rm tot}(t) 
+ N_{\rm tot}(t)/N_0 \right]}
\;.
\label{eq:P}
\end{equation}
Here, the Verhulst factor $N_0$ \cite{VERH1838}
represents an environmental ``carrying capacity'' that might be 
due to limitations on shared resources, such as space, light, or water. 
It prevents the total population from 
indefinite growth, stabilizing it at $O\left( N_0\right)$. 
The interaction matrix $\bf M$ expresses pair interactions between different 
species such that the element $M_{IJ}$ gives the effect of the population 
density of species $J$ on species $I$. Thus, a mutualistic
relationship is represented by both $M_{IJ}$ and $M_{JI}$ being
positive, while both being negative models a competitive relationship. If
they are of opposite signs, we have a predator-prey situation.
To concentrate attention on the effects of interspecies interactions, 
we follow Refs.~\cite{HALL02,CHRI02,COLL03} 
in setting the self-interactions $M_{II} = 0$.
The offdiagonal elements of $M_{IJ}$ are randomly and uniformly distributed 
on $[-1,1]$ as in Ref.~\cite{CHRI02}.
% (but different from Ref.~\cite{HALL02}). 
The interaction matrix is set up at the beginning of the 
simulation and is not changed later, a situation that corresponds to quenched 
disorder in spin-glass models \cite{FISC91}.  
We note that we have not attempted to give $\bf M$ a particularly
biologically realistic form. Some possible modifications 
are discussed in Sec.~\ref{sec:Concl}. 

In this model there is a one-to-one correspondence between genotype (the
$I$th specific bit string) and phenotype (the $I$th row and column of
$\bf M$). Thus, the phenotype specifies both how the $I$th species
influences the other species that are either actually or only
potentially present in the community (the $I$th column) and how it is
influenced by others (the $I$th row). The reproduction
probability $P_I$ provides selection of the ``most fit'' phenotypes
according to these ``traits'' (matrix elements) 
and the populations of the other species present in the community. 
The effect (or lack thereof) of a particular trait depends on the community
in which the species exists, just as a cheetah's superior speed
is only relevant to survival if fast-moving prey is available. 

In each generation, the genomes of the individual offspring organisms are 
subjected to mutation with probability $\mu / L$ per gene and individual. 
Mutated offspring are re-assigned to their new genotypes before the start of 
the next generation. This provides the genetic variability necessary for
evolution to proceed. 

The analytic form of $P_I(\{n_J(t)\})$, Eq.~(\ref{eq:P}),
%, which is the same used in  Refs.~\cite{HALL02,CHRI02}, 
is by no means unique. For instance, one could 
use a ``soft dynamic'' \cite{RIKV02} in which the effects of the interactions 
and the Verhulst factor factorize in $P_I$. Alternatively, 
one could dispense with 
the exponential and use linear or bilinear relationships instead, 
as is most common in population biology \cite{MURR89}. Investigations 
of the effects of such modifications are left for the future.

Our evolution algorithm proceeds in three layers of nested loops.
\begin{itemize}
\item[1.]
Loop over generations $t$.
\item[2.]
Loop over the ${\cal N}(t)$ populated genotypes $I$ in generation $t$.
\item[3a.]
Loop of length $n_I(t)$ over the individuals of genotype $I$. 
Each individual produces 
$F$ offspring for generation $t+1$ with probability 
$P_I(\{n_J(t)\})$, or dies without offspring
with probability $1-P_I$. In either case, no individual survives from
generation $t$ to generation $t+1$. 
\item[3b.]
Loop of length equal to the total number of offspring of genotype $I$ 
generated in loop 3a, 
attempting to mutate each gene of each individual offspring with probability 
$\mu / L$ and moving mutated 
offspring to their respective new genotypes for generation $t+1$.
\end{itemize}

\section{Linear stability analysis}
\label{sec:LSA}

Though neither our model nor the simulations are deterministic, a number of
their gross properties can be understood in terms of a mean-field
approximation that ignores statistical fluctuations and correlations.
The time evolution of the populations is then given by the set of difference
equations, 
\begin{equation}
n_I(t+1)=n_I(t)FP_I(\{n_J(t)\})[1-\mu ]+(\mu
/L)F\sum_{K(I)}n_{K(I)}(t)P_{K(I)}(\{n_J(t)\})+O(\mu ^2)\;,  
\label{eq:EOM}
\end{equation}
where $\sum_{K(I)}$ runs over the $L$ genotypes $K(I)$ that differ from $I$ by
one single mutation (i.e., the Hamming distance \cite{HAMM86} $H_{KI} =1$). 
The corrections of $O(\mu^2)$ correspond to multiple mutations in 
single individuals.
Naturally, a full investigation of this equation is highly
non-trivial, even in the absence of random noise.  
In particular, Eq.~(\ref{eq:EOM})
can be regarded as a logistic map in a $2^L$-dimensional space. Logistic
maps are known to admit, in general, both fixed points and cycles of
non-trivial periods \cite{MURR89}. To keep both analysis of simulations and
theoretical investigations manageable, we choose parameters in such a
way that we can focus our attention on fixed points. 
%A fixed point with 
%{\em non-negative} values of $n_I$ 
%may be associated with a physical population
%and will be termed a ``feasible fixed-point community'' \cite{ROBE74}. 

Of course, what we simulate are stochastic processes. However, as long as
the mutation rate is well below the error threshold for mutational melt-down 
\cite{DROS01,EIGE71,EIGE88,BAAK99,HALL02,COLL03}, 
new successful mutations can become
fixed in the population before another successful mutant arises. This results 
in a separation of time scales between the ecological scale of fluctuations
within fixed-point communities and the evolutional scale of the
durations of such communities \cite{DOEB00}, which are 
known as quasi-steady states
(QSS) or quasi evolutionarily stable strategies (q-ESS) \cite{HALL02,CHRI02}. 
By ignoring mutations, we can make some analytical progress and gain some
insight into the nature and stability properties of these QSS.

\subsection{$\cal N$-species fixed points}
\label{sec:Nfp}

A fixed point of Eq.~(\ref{eq:EOM}), defined by 
\begin{equation}
n_I^{*}(t+1)=n_I^{*}(t)\equiv n_I^{*}\,\,,  
\label{eq:fp}
\end{equation}
is characterized by having only $\cal N$ ($\leq 2^L$) non-vanishing $n_I^{*}$. 
(Henceforth, an asterisk will signify a quantity as a fixed-point value.)
Following Ref.~\cite{ROBE74}, we denote a fixed point as {\it feasible\/} 
if $n_I^* > 0$ for all $\cal N$ values of $I$. 
Corresponding to the coexistence of $\cal N$ genotypes, such a point will
be simply referred to as an ``$\cal N$-species fixed point.'' 
When mutations are ignored, Eq.~(\ref{eq:EOM}) reduces to 
\begin{equation}
n_I(t+1)=n_I(t)FP_I(\{n_J(t)\})\,\,.  
\label{eq:mu=0}
\end{equation}
Specializing to the form of $P_I$ given by Eq.~(\ref{eq:P}), we see that all
single-species fixed points are trivially ``identical,'' with 
$n_I^{*}=N_{\mathrm{tot}}^{*}=N_0\ln (F-1)$. 
This somewhat unrealistic result is just a consequence of our choice of
$M_{II} = 0$ and can be avoided by lifting this restriction. However, 
the absence of 
self-interactions places no restriction on the main purpose of our work
-- the exploration of the effects of random but time independent 
interspecies interactions.

Proceeding to the ${\cal N} \geq 2$ cases, the existence of an 
$\cal N$-species fixed
point depends critically on the submatrix $\tilde{\mathbf{M}}$, with matrix
elements $M_{IJ}$ in which both $I$ and $J$ are among the 
$\cal N$ species in question. 
(We shall use the tilde to emphasize that a quantity corresponds to an 
$\cal N$-species subspace, rather than to the full, $2^L$-dimensional 
genotype space.)
In particular, if $\tilde{\mathbf{M}}$ is non-singular, then a
unique fixed point exists, as we show below. On the other hand,
singular $\tilde{\mathbf{M}}$'s may result in a variety of ``degenerate''
cases. We provide just two examples to illustrate the mathematical richness of 
singular $\tilde{\mathbf{M}}$'s. 
If all elements vanish, then the behavior in this
subspace is highly degenerate, with the \emph{total} population given again
by $N_{\mathrm{tot}}^{*}=N_0\ln (F-1)$, regardless of $\cal N$. However, the
fractions of each species, 
\begin{equation}
\rho_I^*\equiv n_I^{*}/N_{\mathrm{tot}}^{*}
\;,
\label{eq:deff} 
\end{equation}
are completely undetermined, an understandable consequence of having
dynamically indistinguishable species. Another example of a singular 
interaction submatrix is 
$\tilde{\mathbf{M}} = 
\left( 
\begin{array}{cc}
0 & m>0 \\ 
0 & 0
\end{array}
\right) $ 
in an ${\cal N}=2$ subspace. 
Then, Eq.~(\ref{eq:mu=0}) will drive one of
the species to extinction, so that no fixed point with both $n_I > 0$ 
(stable or unstable) can exist in this two-dimensional subspace, and the
system collapses to ${\cal N} =1$. 
For the remainder of this article, we shall study analytically
only the dynamical behavior of 
interacting species with nonsingular $\tilde{\mathbf{M}}$'s.

To proceed, we insert Eq.~(\ref{eq:fp}) into Eq.~(\ref{eq:mu=0}) and
%,
%invoking $n_I^{*}>0$, 
arrive at the $\cal N$ equations 
\begin{equation}
FP_I(\{n_J^{*}\})=1\,\,.  
\label{eq:fix1}
\end{equation}
With our choice of $P_I$, these lead to 
\begin{equation}
\sum_JM_{IJ}\rho_J^*=\frac{N_{\mathrm{tot}}^{*}}{N_0}-\ln (F-1)\;.
\label{eq:fix2}
\end{equation}
(The analysis in this section remains valid even if $M_{II} \neq 0$,
although $M_{II} = 0$ is the case explicitly considered elsewhere
in this paper.) Note that the right-hand side of Eq.~(\ref{eq:fix2}) 
is \emph{independent} of $I$. Since $\tilde{\mathbf{M}}$ is non-singular, 
we define its inverse by $\tilde{\mathbf{W}}$, with elements $\tilde{W}_{IJ}$: 
\begin{equation}
\tilde{W}_{IJ}\equiv (\tilde{\mathbf{M}}^{-1})_{IJ}\,\,.
\label{eq:W}
\end{equation}
Then Eq.~(\ref{eq:fix2}) can be inverted to give 
\begin{equation}
\rho_I^*=\left[ \frac{N_{\mathrm{tot}}^{*}}{N_0}-\ln (F-1)\right]
\sum_J\tilde{W}_{IJ}
\,\,.  
\label{eq:fix3}
\end{equation}
The only unknown in this equation, $N_{\mathrm{tot}}^{*}$, can now be found
via the normalization condition, $\sum \rho_I^*=1$, which leads to 
\begin{equation}
N_{\mathrm{tot}}^{*}/N_0=\ln (F-1)+1/\tilde{\Sigma}\;,  
\label{eq:fix4}
\end{equation}
where 
\begin{equation}
\tilde{\Sigma}\equiv \sum_{IJ}\tilde{W}_{IJ}\,\,.
\label{eq:Sigma}
\end{equation}
Putting these into Eq.~(\ref{eq:fix3}), we have the explicit form of the
fixed-point populations: 
\begin{equation}
n_I^{*} 
=
N_0 \left[ \ln (F-1)+ \frac{1}{\tilde{\Sigma}} \right]
\frac{\sum_J\tilde{W}_{IJ}}{\tilde{\Sigma}}
\;.
\label{eq:fix5}
\end{equation}

Although Eq.~(\ref{eq:fix5}) appears to provide fixed-point values for 
\emph{any} choice of control parameters, we emphasize that it is
applicable only for a limited range of $F$ and $\tilde{\mathbf{M}}$. 
The subtlety lies in its stability properties within the 
$\cal N$-species subspace. 
First, we remind the reader that, even in the case of ${\cal N}=1$, 
there are a variety of behaviors. Solutions may have a stable fixed point
with either monotonic or oscillatory decay of small deviations, 
or they may show bifurcations,
period doubling, and chaos \cite{MURR89}. In the present 
study, we are interested
in the effects of the interspecies interactions, and we therefore choose to
focus on systems with monotonically decaying fluctuations in the
noninteracting limit. This means 
that small fluctuations about the fixed point must decay as 
$\delta(t+1)/\delta(t) \in \langle 0,1 \rangle$. For the 
single-species and ${\mathbf M} = {\mathbf 0}$ cases, 
this restriction is easily translated to the condition $2<F \alt 4.5$ for
the fecundity. We use $F=4$ in all of our simulations.

Next, through a straightforward linear stability analysis around the 
$\cal N$-species fixed point, we obtain a condition that represents
decay of all small perturbations, i.e., all the eigenvalues of
the matrix 
\begin{equation}
\left. \frac{\partial n_I(t+1)}{\partial n_J(t)} \right|_{\{n_I^*\}}
\label{eq:A1}
\end{equation}
must have real parts, 
lying between $-1$ and 1. Carrying out the differentiation
and using Eq.~(\ref{eq:mu=0}) to simplify the result, 
we find this matrix to be of the form
${\bf 1} + \tilde{\bf \Lambda}$, 
%$\delta_{IJ} - \tilde{\Lambda}_{IJ}$, 
where $\bf 1$ is the $\cal N$-dimensional unit matrix and 
the elements of $\tilde{\bf \Lambda}$ are given by
\begin{equation}
\tilde{\Lambda}_{IJ}  
= 
\left( 1 - \frac{1}{F} \right)
\rho_I^* \left( M_{IJ} - \ln (F-1) - 2/\tilde{\Sigma} \right)
\;.
\label{eq:A2}
\end{equation}
Our criterion translates to the requirement that 
all the eigenvalues of $\tilde{\bf \Lambda}$ 
must have real parts that lie in $\langle -2,0 \rangle$. 
Fixed points with this property will be called ``internally stable.''
Unfortunately, this criterion cannot be made more explicit. Given $F$
and a set of $M_{IJ}$, $\tilde{\bf \Lambda}$ must be constructed using 
Eqs.~(\ref{eq:W})--(\ref{eq:Sigma}) and diagonalized. 
The matrix $\tilde{\bf \Lambda}$ is recognized as what is known in ecology 
as the {\it community matrix\/} \cite{MURR89} of the $\cal N$-species fixed 
point for the discrete-time dynamic defined by Eq.~(\ref{eq:mu=0}). 

It has been shown, both numerically \cite{GARD70} and analytically 
\cite{MAY72}, that the proportion of large, random matrices 
for which all eigenvalues have a negative real part 
vanishes as the proportion of nonzero matrix
elements increases. This has been used as an argument that highly
connected ecosystems are intrinsically unstable \cite{MAY72}, contrary
to ecological intuition.  
However, the matrix that must be studied to determine the internal
stability of a fixed
point is the community matrix $\tilde{\bf \Lambda}$, which has a complicated 
relationship to the interaction matrix $\bf M$ and should not be expected to 
have a simple element distribution. 
In fact, for some bilinear population dynamics models 
there is numerical evidence that most {\it feasible\/} fixed points 
are also internally stable \cite{ROBE74,GILP75}. However, 
the relations between connectivity and stability have not yet been fully
clarified and are still being discussed \cite{ROZD01,WILM02}. 

Of course, issues of internal stability are somewhat academic for
simulations. In practice an internally unstable fixed point could not be
observed for more than a brief time, especially since the populations would
be driven away from such fixed points by the noise due to both the birth/death
process and the mutations.

\subsection{Stability against other species}
\label{sec:Ext}

Since mutations are essential for the long-time evolutionary behavior, the
``external stability'' properties of the fixed-point communities are 
important. Even if the population corresponds to 
an internally stable $\cal N$-species fixed point,
mutations will generate small populations of ``invader'' species, i.e.,
genotypes outside the resident 
$\cal N$-species community. Denoting such invaders by the
subscript $i$ and linearizing Eq.~(\ref{eq:mu=0}) about the 
$\cal N$-species fixed point, we
see that the important quantity is the 
multiplication rate of the invader species in the limit of vanishing 
$n_i/N_{\rm tot}$,
\begin{equation}
\frac{n_i(t+1)}{n_i(t)}
=
\frac F{1+(F-1)\exp \left[ \left( 1-\sum_{JK}M_{iJ}{\tilde W}_{JK}\right) 
/ \tilde{\Sigma} \right] }\;.
\label{eq:other}
\end{equation}
To obtain this result we exploited Eqs.~(\ref{eq:fix4}) and~(\ref{eq:fix5}). 
Explicitly, the condition for stability against the invader, 
$n_i(t+1)/n_i(t) < 1$, reduces to the requirement that the argument of the
exponential function in Eq.~(\ref{eq:other}) must be positive. 
We also note that, if $M_{iJ}=0$ for all $J$ in the resident community, 
then the multiplication rate of the invader equals unity.  
In population biology the Lyapunov exponent 
$\ln \left[ n_i(t+1)/n_i(t) \right]$ is known as the {\it invasion
fitness\/} of the mutant with respect to the resident community 
\cite{DOEB00,METZ92}. 
From Eq.~(\ref{eq:other}) it becomes clear that the success of an invader 
depends not only on its direct interactions $M_{iJ}$ with each of the resident 
species, but also on the interactions between the resident species through 
the inverse interaction submatrix ${\tilde W}_{JK}$. 

\section{Simulation results}
\label{sec:MC}

The model described in Sec.~\ref{sec:Mod} was studied by Monte Carlo
simulations with the following parameters: $L=13$, $F=4$, $N_0 =2000$, 
and $\mu = 10^{-3}$. The random matrix $\bf M$ (with zero diagonal and other 
elements randomly distributed on $[-1,1]$) was chosen at the beginning of 
each simulation run and then kept constant for the duration of the run. 
In this regime both the number of populated
species ${\cal N}(t)$ and the total population 
$N_{\rm tot}(t) \sim N_0 \ln(F-1) \approx 2200$ are substantially
smaller than the number of possible species, 
${\cal N}_{\rm max} = 2^L = 8192$.
This appears biologically reasonable in view of the enormous number of 
different possible genotypes in nature. 
(But see further discussion in Sec.\ref{sec:Stat} and 
Appendix~\ref{sec:AC}.) In a QSS, the average
number of mutant offspring of species $I$ in generation $t+1$ 
is approximately $\mu n_I(t)$. Thus, with the mutation rate used here,
each of the dominant species will produce of the order of one mutant
organism per generation. As shown in Sec.~\ref{sec:Stab}, during a QSS
most of these mutants become extinct after one generation. However, due to
the small genome size, the same mutant of a species with a large
population will be re-generated repeatedly by mutation from the parent
species. 

A high-population genotype with its ``cloud'' of closely related
low-population mutants could be 
considered a quasi-species in the sense of Eigen, with the high-population 
genotype as the ``wildtype'' \cite{DROS01,COLL03,EIGE71,EIGE88,BAAK99}. 
An alternative interpretation of the model is therefore as one of the 
coevolution of quasi-species \cite{COLL03}, in which the successful invasion 
of a resident community by a mutant represents a speciation event in the 
lineage of the parent genotype. 

\subsection{General features}
\label{sec:Gen}

Most of our simulations were started with a small population of 100
individuals of a randomly selected 
single genotype, corresponding to the entry into an
empty ecological niche by a small group of identical individuals. 
However, runs starting from a random number of small populations of
different species give essentially the same results. Generally, the
initial species are not likely to be stable against mutants, and 
they usually become extinct within 100 generations. 
To eliminate any short initial transients, 
most of the quantitative analyses presented here are 
based on time series from which the first 4096 generations were removed. 
The simulated quantities were recorded every 16 generations. This time
resolution was chosen to be just larger than the average time it would 
take for the descendants of a single individual of a mutant genotype $i$ to
completely replace a resident genotype $J$ of $N_0=2000$
individuals in the case
of a maximally aggressive mutant, $M_{iJ}=+1$ and $M_{Ji}=-1$. This time,
which is obtained by numerical solution of Eq.~(\ref{eq:other}), is about
15~generations and represents a ``minimum growth time" for the model.
Sampling at shorter intervals would mostly add random noise to the
results, while sampling on a much coarser scale could miss important
evolutionary events. It is easy to show, by solving Eq.~(\ref{eq:other})
analytically for short times, that the growth time and thus the optimal
sampling interval increases logarithmically with $N_0$.   
Most quantitative results in this paper are based on averages over 16 
independent 
simulation runs of $2^{25} = 33\,554\,432$ generations each \cite{COMP}. 

We define the {\it diversity\/} of the population as the number of species 
with significant populations, thus excluding small populations of mainly
unsuccessful mutants of the major genotypes. Operationally we define the 
diversity as $D(t) = \exp \left[S \left( \{ n_I(t) \} \right) \right]$, 
where $S$ is the information-theoretical entropy \cite{SHAN48,SHAN49},
\begin{equation}
S\left( \{ n_I(t) \} \right) 
=
- \sum_{\{I | \rho_I(t) > 0 \}} \rho_I(t) \ln \rho_I(t)
\label{eq:S}
\end{equation}
with
\begin{equation}
\rho_I(t) = n_I(t) / N_{\rm tot}(t)
\;.
\label{eq:rho}
\end{equation}
This measure of diversity is known in ecology as the Shannon-Wiener
index \cite{KREB89}. 
It is different from the definition of diversity as the number of
populated species (known in ecology as the {\it species richness\/}
\cite{KREB89}) that was used in Refs.~\cite{HALL02,CHRI02}. The entropy-based 
measure significantly reduces the noise during QSS, caused by
unsuccessful mutations.

The Shannon-Wiener diversity index
is shown in Fig.~\ref{fig:timeser}(a) for a simulation
of $10^6$ generations, together with the 
normalized total population, $N_{\rm tot}(t)/[N_0 \ln(F-1)]$. We see quiet
periods during which $D(t)$ is constant except for small fluctuations,
separated by periods during which it fluctuates wildly. 
The total population $N_{\rm tot}(t)$ 
is enhanced relative to its noninteracting fixed-point value during the 
quiet periods, while it decreases toward the vicinity of this value 
during the active periods. 

To verify that different quiet periods indeed correspond to different
resident communities 
$\{ n_I \}$, we show in Fig.~\ref{fig:spec} the genotype labels $I$
(integers between 0 and $2^L-1$, corresponding to the decimal
representation of the genotype bit string) versus time
for the populated species. The populations of the different species are
indicated by the grayscale (by color online). We see that, in general,
the community is completely rearranged during the active periods, so that
the quiet periods can be identified with the QSS of the ecology. 
An exception is afforded by the event near $t = 1.5 \times
10^5$~generations in Figs.~\ref{fig:timeser}(a) 
and~\ref{fig:spec}(a). In this instance the populations of the dominant
species decrease, and the total population is for a brief time spread over
a larger number of species. However, the dominant species ``regain their
footing," and the original resident 
community continues for approximately another 
50\,000 generations. This situation is reminiscent of rare events in
nucleation theory \cite{RIKV94}, 
in which a fluctuation of the order of a critical or
even a supercritical droplet nevertheless may decay back to the metastable
state.

\subsection{Stability of communities against invaders}
\label{sec:Stab}

To investigate the stability properties  
of individual communities against invaders, we
chose from a particular simulation run of $10^6$
generations ten different QSS of durations longer than 20\,000 generations.
The genotype labels and fixed-point 
populations that characterize these QSS are given in the
first four columns of Table~\ref{table1}.  
The population-weighted average of the Hamming distances $H_{IJ}$ 
between pairs of genotypes in the community,
\begin{equation}
\langle {H} \rangle
=
\sum_{I} \sum_{J>I} \rho_I \rho_J 
\left( \frac{1}{1 - \rho_I} + \frac{1}{1 - \rho_J} \right) H_{IJ} 
\;,
\label{eq:avHamm}
\end{equation}
and the corresponding standard deviation, 
\begin{equation}
\sigma_{H}
=
\sqrt{
\frac{\cal N}{{\cal N} -1}
\sum_{I} \sum_{J>I} \rho_I \rho_J 
\left( \frac{1}{1 - \rho_I} + \frac{1}{1 - \rho_J} \right) 
\left( H_{IJ} - \langle H \rangle \right)^2
} ,
\label{eq:devHamm}
\end{equation}
are shown in the fifth and sixth column, respectively. 
Even though the community as a whole moves far through the 
$2^L$-dimensional population space, the different QSS communities 
are seen to retain the property that they consist 
of relatively close relatives
(which of course are all descendants of the single, initial genotype). 
In Sec.~\ref{sec:Stat} we demonstrate that $\langle H \rangle$
and $\sigma_H$ remain in the range shown in Table~\ref{table1}, even
during very long simulations.  

The averages and 
standard deviations of the offdiagonal interaction-matrix elements
${M}_{IJ}$ between the community members are shown in the seventh and
eighth column of Table~\ref{table1}, respectively. 
They show that the QSS communities are
strongly mutualistic, as was also observed for the stable states in 
Ref.~\cite{KAUF91}. In contrast, the matrix elements of 
22 feasible, but otherwise randomly chosen, communities \cite{FEAS,TREG79} 
were found to be approximately uniformly distributed on $[-1,1]$. 

\begin{figure}[ht] 
\includegraphics[angle=0,width=.47\textwidth]{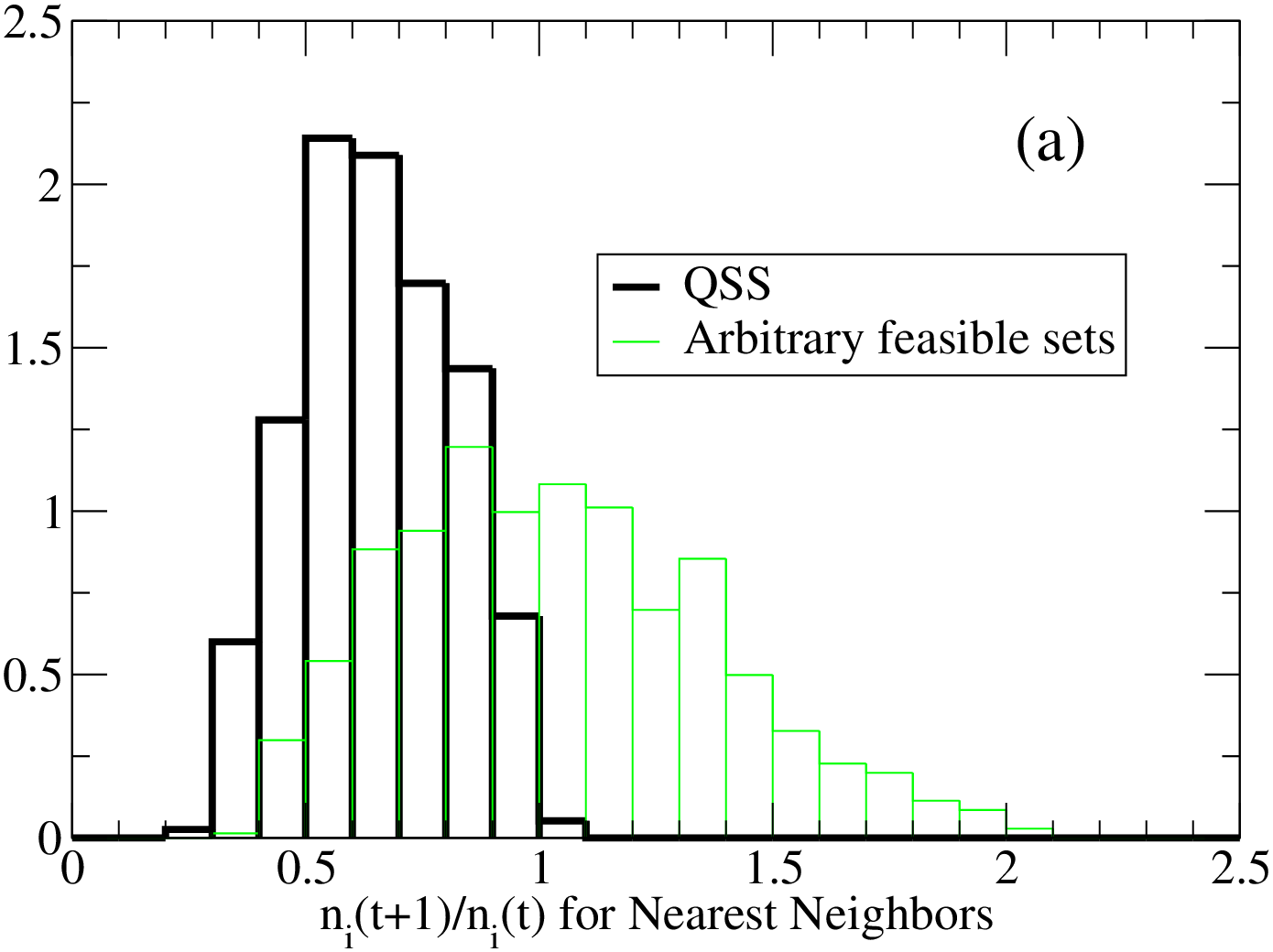} 
\hspace{0.5truecm}
\includegraphics[angle=0,width=.47\textwidth]{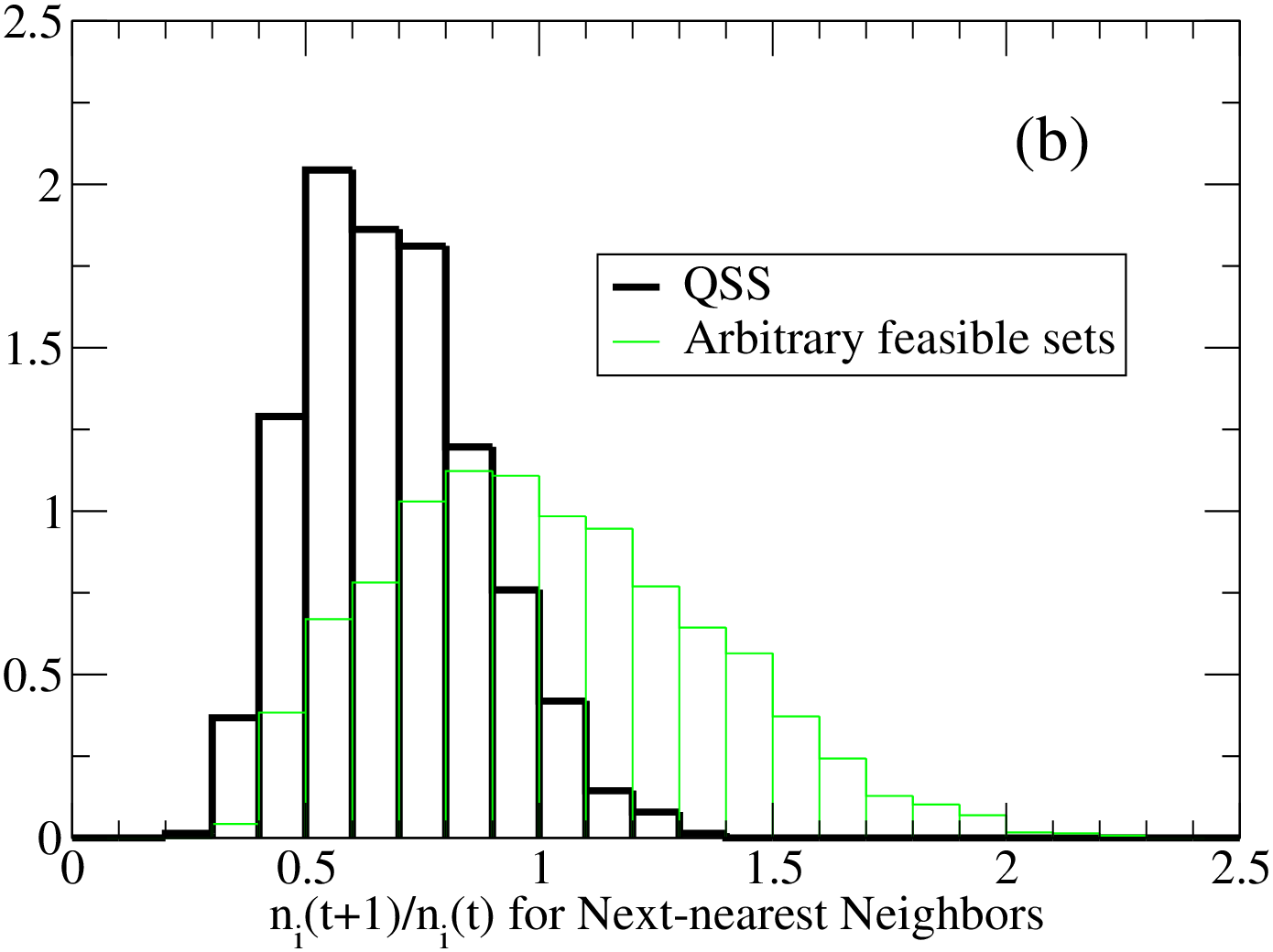} 
\caption[]{
(Color online.) 
Histograms of the multiplication rate (exponential of the invasion fitness)
for mutant species $i$ against each of ten specific
QSS (thick, black lines), compared with the same quantity against randomly
chosen, feasible communities [thin, gray lines (green on
line)]. The multiplication rates are calculated from Eq.~(\ref{eq:other}). 
(a)
When the mutant species differ from the resident community by a Hamming
distance of one (nearest neighbors \protect\cite{ENDNOTE}). 
(b)
When the mutant species differ from the resident community by a Hamming
distance of two (next-nearest neighbors). 
See discussion in Sec.~\protect\ref{sec:Stab}. 
}
\label{fig:eigenval}
\end{figure}

In Fig.~\ref{fig:eigenval} we show histograms of the multiplication
rates Eq.~(\ref{eq:other}) (i.e., the exponential 
function of the invasion fitness) of mutants that differ from
the resident species by one single mutation (nearest-neighbor species,
$\min_J H_{iJ}=1$), 
and by two mutations (next-nearest-neighbor species, 
$\min_J H_{iJ}=2$) 
\cite{ENDNOTE}. 
Only a very small proportion of the nearest-neighbor mutants have a 
multiplication rate above unity [Fig.~\ref{fig:eigenval}(a)]. 
(In our sample, this very small proportion came from just one of the
ten QSS considered.) Among the next-nearest neighbors, on the other hand, 
a not insignificant proportion may be successful invaders
[Fig.~\ref{fig:eigenval}(b)]. The picture for third-nearest neighbors
(not shown) is essentially the same as for the next-nearest neighbors. 
We also considered the multiplication rates for neighbors of 
the two most long-lived QSS 
observed in our simulations, which lasted about $1.0 \times 10^7$ and 
$1.4 \times 10^7$ generations, respectively. They both had no nearest neighbors 
with multiplication rates above unity, and the proportions for 
second-and third-neighbors were significantly smaller than those included in
Fig.~\ref{fig:eigenval}.
In both parts of Fig.~\ref{fig:eigenval} are also shown the
multiplication rates for neighbors of the 22 randomly chosen, feasible
communities introduced in the previous paragraph. 
In that case, approximately half of the neighbors at any Hamming distance
have multiplication rates above unity. 

Summarizing the results from 
Table~\ref{table1} and Fig.~\ref{fig:eigenval}, we see that a long-lived
QSS community is characterized by strongly mutualistic interactions and is
surrounded by a ``protection zone'' of closely related 
genotypes that are very unlikely to successfully invade the resident community. 
The evidence from the two most long-lived QSS indicates that the
average lifetime of a QSS is positively correlated with the extent of
the protection zone.
By contrast, a randomly chosen, feasible community has an
approximately uniform distribution of ${M}_{IJ}$ over $[-1,1]$. 
It also typically has no protection zone and so  
is more vulnerable to invasion than a QSS. 
It is clear from the fact that none of our randomly chosen feasible communities 
turned out to be a QSS that QSS are relatively 
rare in this model, even among feasible communities. We believe this 
may be a favorable condition
for continuing evolution, as the ecology can move rapidly from QSS to QSS
through series of unstable communities.

\subsection{QSS lifetime statistics}
\label{sec:Life}

The protection zone surrounding a QSS community acts much like the free-energy
barrier that separates a metastable state in a physical system from the
stable state or other metastable states. In both cases a sequence of
improbable mutations or fluctuations 
is required to reach a critical state that 
will lead to major rearrangement of the system. It is thus natural to
investigate in detail the lifetime 
statistics of individual QSS in the way common in the
study of metastable decay \cite{RIKV94,RIKV94A}. 
To this effect we started simulations from each of the ten QSS
communities discussed in Sec.~\ref{sec:Stab} 
and ran until the overlap function with the initial community, 
\begin{equation}
{\cal O}(0,t) 
= 
\frac{\sum_I n_I(0) n_I(t)}{\sqrt{\sum_I n_I(0)^2 \sum_J n_J(t)^2}}
\;,
\label{eq:over}
\end{equation}
became less than 0.5, at which time the system was 
considered to have escaped from the initial state. 
(We note that the overlap function defined here is simply the cosine of the 
angle between two unit vectors in the space of population vectors $\{ n_I \}$.)
The precise value of the 
cutoff used is not important, as long as it is low enough
to exclude fluctuations during the QSS \cite{RIKV94A}. 
The composition of the population at this time (the ``exit 
community'') was then recorded and the run terminated. 
Each individual initial QSS was simulated for 300 independent 
escapes. The means and standard deviations of the individual lifetime
distributions are shown in the last two columns of Table~\ref{table1}. 
The mean lifetimes were found to range over
about one order of magnitude, from approximately 10\,000 to 80\,000 generations.
In most cases the individual lifetime distributions were exponential
(for which the standard deviation equals the mean), or they had an
exponential component that described the behavior in the short-time end
of the range of observed lifetimes. In about half of the QSS studied, a
tail of very long lifetimes beyond the exponential part (indicated by
a standard deviation significantly larger than the mean) was also
observed. We further found
that an initial QSS does not always escape via the same exit
community. Typically, the genotypes in the exit community that are
not present in the initial QSS differ from those in 
the original community by a Hamming distance of two or three. 
This picture is quite consistent with the stability properties of the QSS 
discussed in Sec.~\ref{sec:Stab}. 

\begin{figure}[ht] 
\includegraphics[angle=0,width=.47\textwidth]{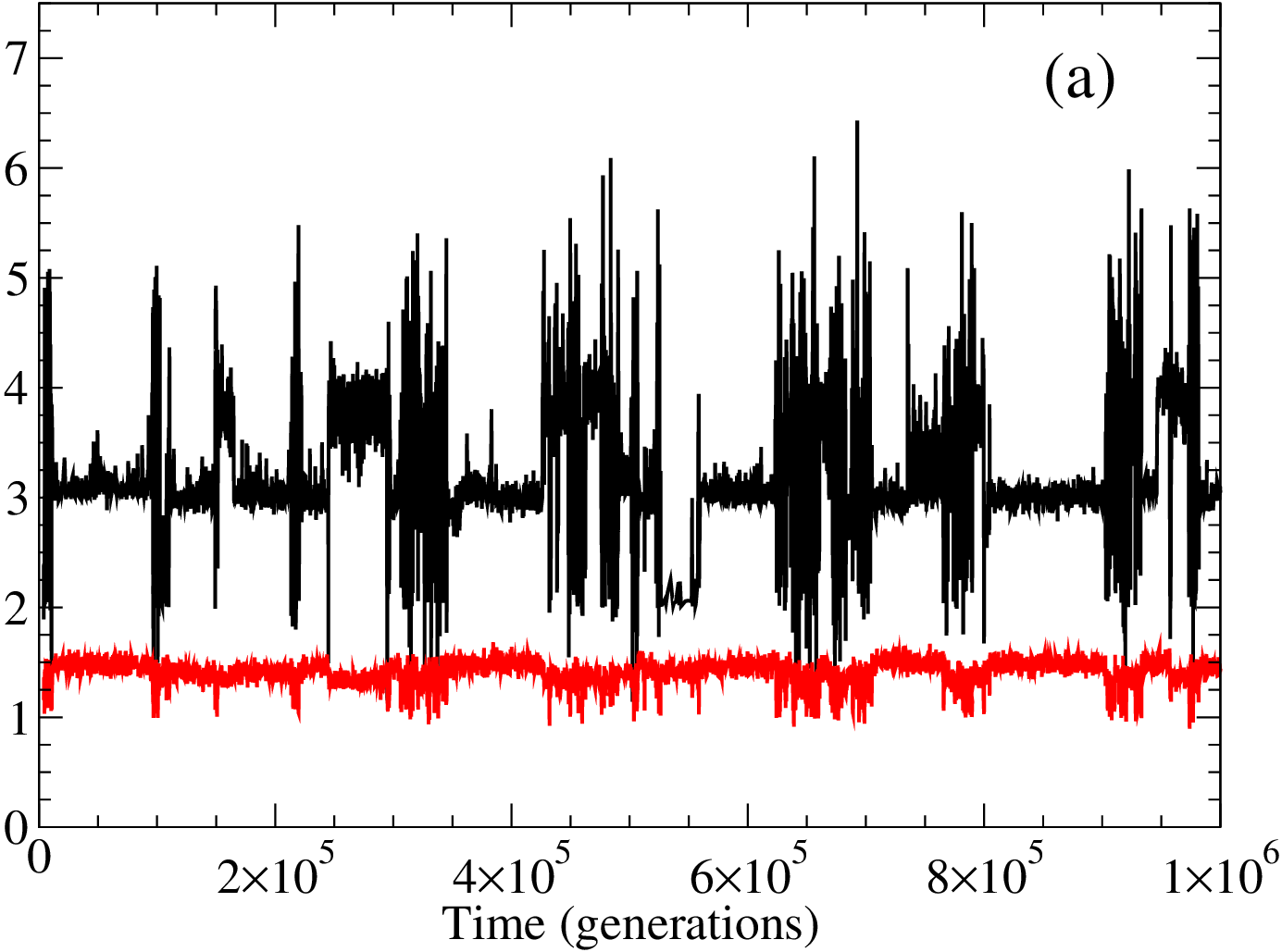} 
\hspace{0.5truecm}
\includegraphics[angle=0,width=.47\textwidth]{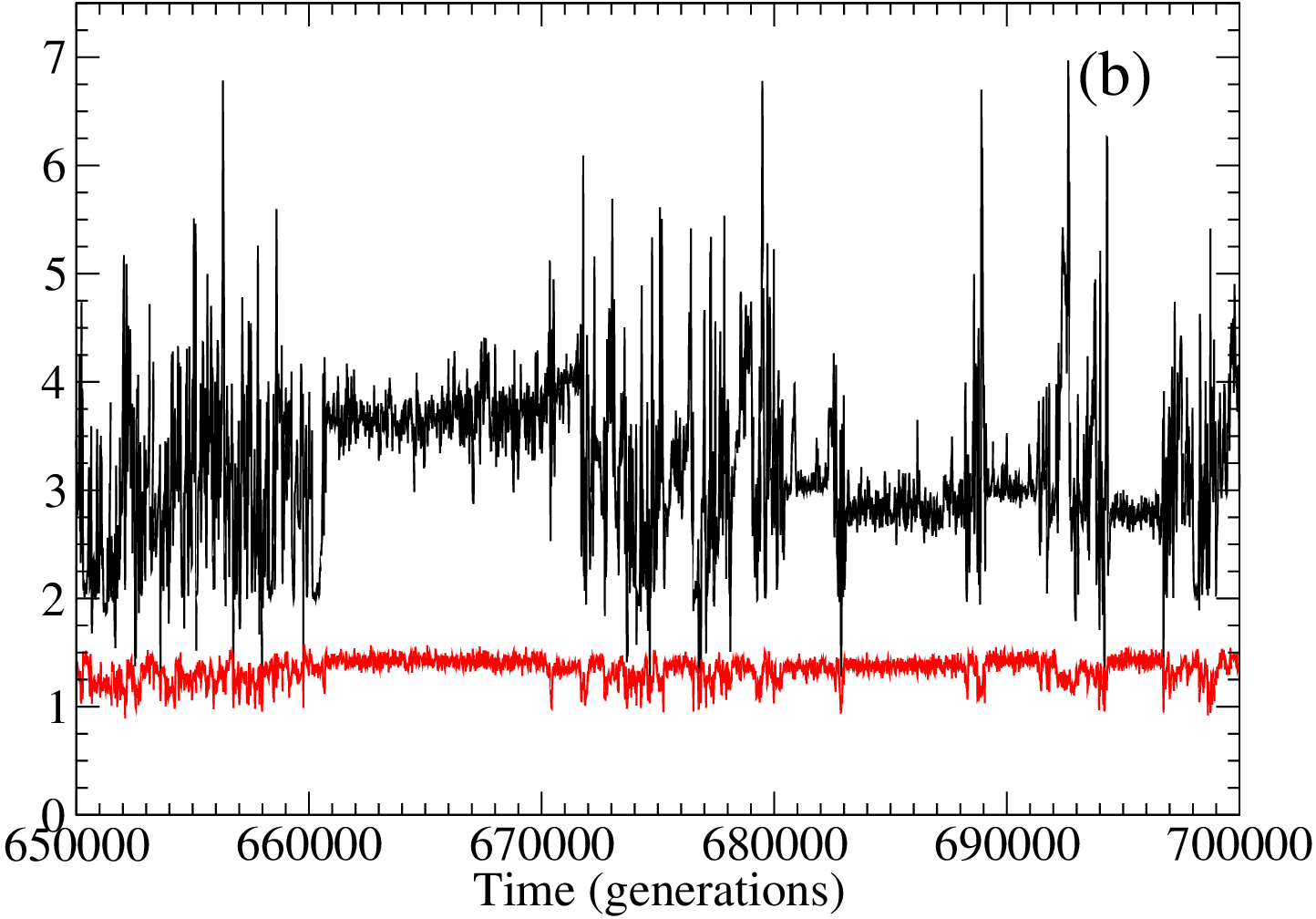} 
\caption[]{
(Color online.) 
Time series from a simulation of $10^6$ generations.
The model parameters, which are the same in the subsequent 
figures, are: mutation rate $\mu = 10^{-3}$ per individual,
carrying capacity $N_0 = 2000$, fecundity $F=4$, and genome length $L=13$. 
Top curve (black): Shannon-Wiener diversity index $D(t) = \exp[S(\{n_I(t)\})]$. 
Bottom curve (gray, red online): 
normalized total population $N_{\rm tot}(t)/[N_0 \ln (F-1)]$. 
(a) 
The whole time series, showing intervals of quasi-steady states (QSS)
separated by periods of high activity. 
(b)
Part of one of the periods of high activity, 
shown on a time scale expanded twenty
times. Several shorter QSS are resolved between bursts of
high activity. Comparison with (a) suggests statistical self-similarity. 
See discussion in Secs.~\protect\ref{sec:Gen} and~\protect\ref{sec:Life}. 
}
\label{fig:timeser}
\end{figure}

\begin{figure}[ht] 
\includegraphics[angle=0,width=.47\textwidth]{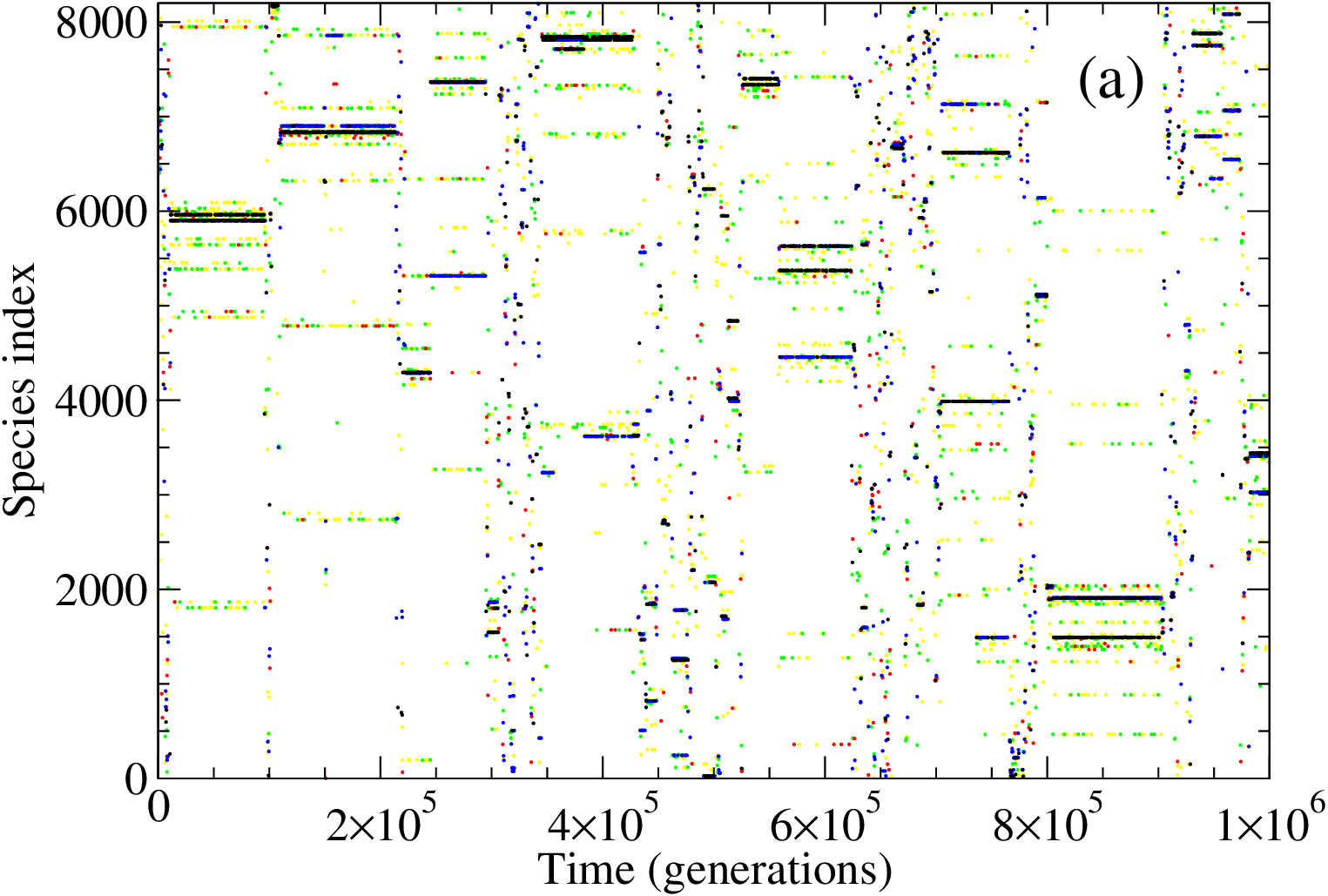} 
\hspace{0.5truecm}
\includegraphics[angle=0,width=.47\textwidth]{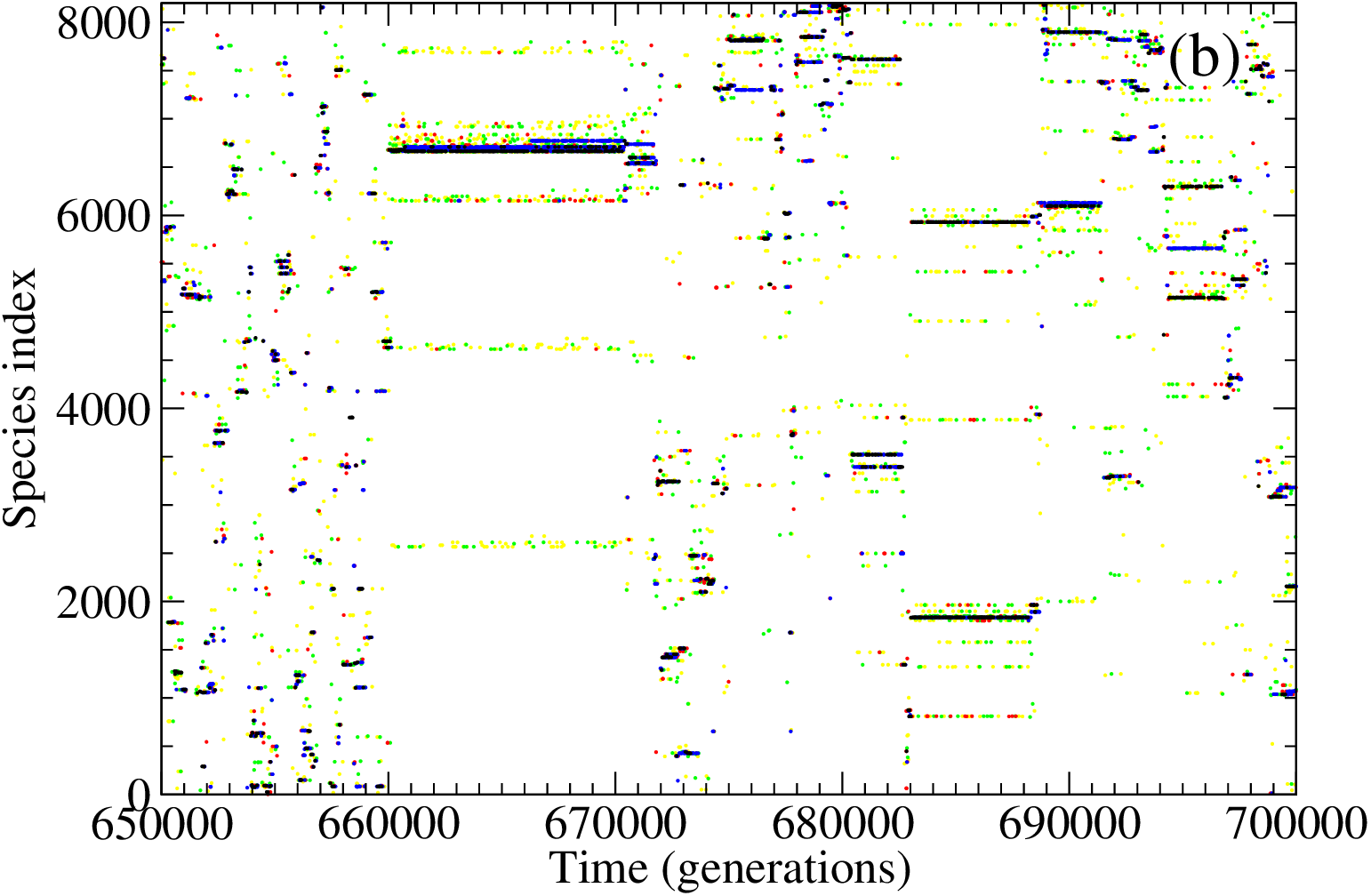} 
\caption[]{
(Color online.) 
Genotype label $I$ versus time for the same simulation run shown in
Fig.~\ref{fig:timeser}. In order of decreasing darkness from black to very
light gray the symbols indicate $n_I \ge 1001$, $n_I \in [101,1000]$, 
$n_I \in [11,100]$, $n_i \in [2,10]$, and $n_I=1$. (Online the colors are
black, blue, red, green, and yellow in the same order.) 
Note that the difference between the label for two species bears no simple
relation to their Hamming distance. 
Each QSS is composed of a different set of
species, punctuated by periods during which the population moves
rapidly through genotype space. 
(a)
Corresponding to Fig.~\ref{fig:timeser}(a), sampled every 
%320 
960 generations 
to facilitate plotting. 
(b)
Corresponding to Fig.~\ref{fig:timeser}(b), sampled every 
%16 
48 generations. 
See discussion in Secs.~\protect\ref{sec:Gen} and~\protect\ref{sec:Life}. 
}
\label{fig:spec}
\end{figure}

The range of the lifetimes of the ten QSS discussed above was limited:
for a period to be identified by eye 
as a QSS, its lifetime could not be too short,
while the lifetimes were bounded above by the length of the simulation run. 
The variation of a factor of ten within these
limits indicates that the dynamical behavior 
of the system may display a very wide
range of time scales. Another indication to this effect is provided by
the details of the activity within periods that appear as
high-activity in Figs.~\ref{fig:timeser}(a) and~\ref{fig:spec}(a).  
Such detail is shown in Figs.~\ref{fig:timeser}(b) and~\ref{fig:spec}(b).
Statistically, the picture in these expanded figures 
is similar to the one seen on the larger scale, 
with shorter quiet periods punctuated by
even shorter bursts of activity. These observations, which are similar to
Refs.~\cite{HALL02,CHRI02}, suggest statistical 
self-similarity at least over some range
of time scales. Indications of self similarity have also been 
seen in fossil diversity records \cite{SOLE97}. 

\begin{figure}[ht] 
\vspace{0.5truecm}
\includegraphics[angle=0,width=.47\textwidth]{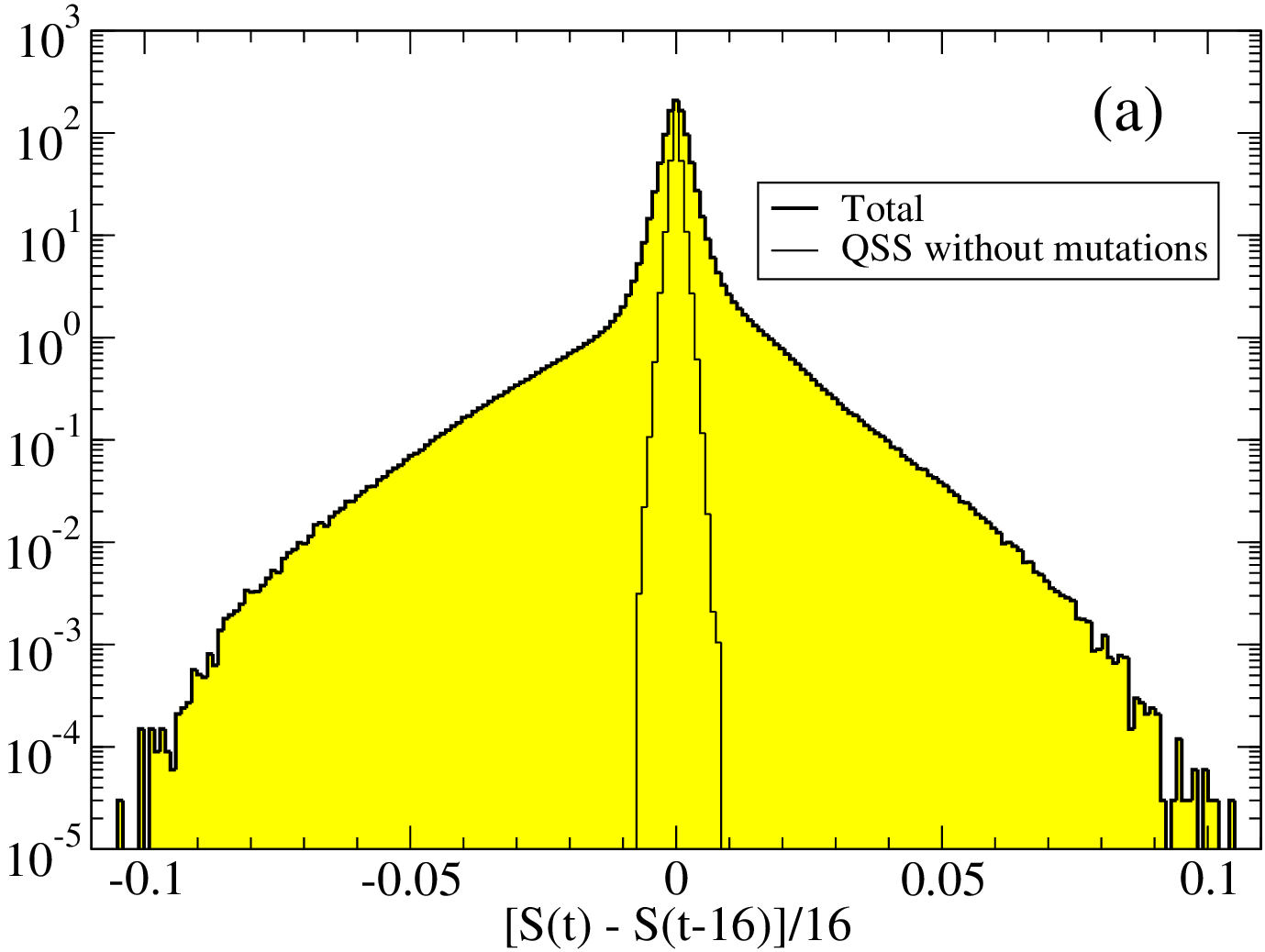} 
\hspace{0.5truecm}
\includegraphics[angle=0,width=.47\textwidth]{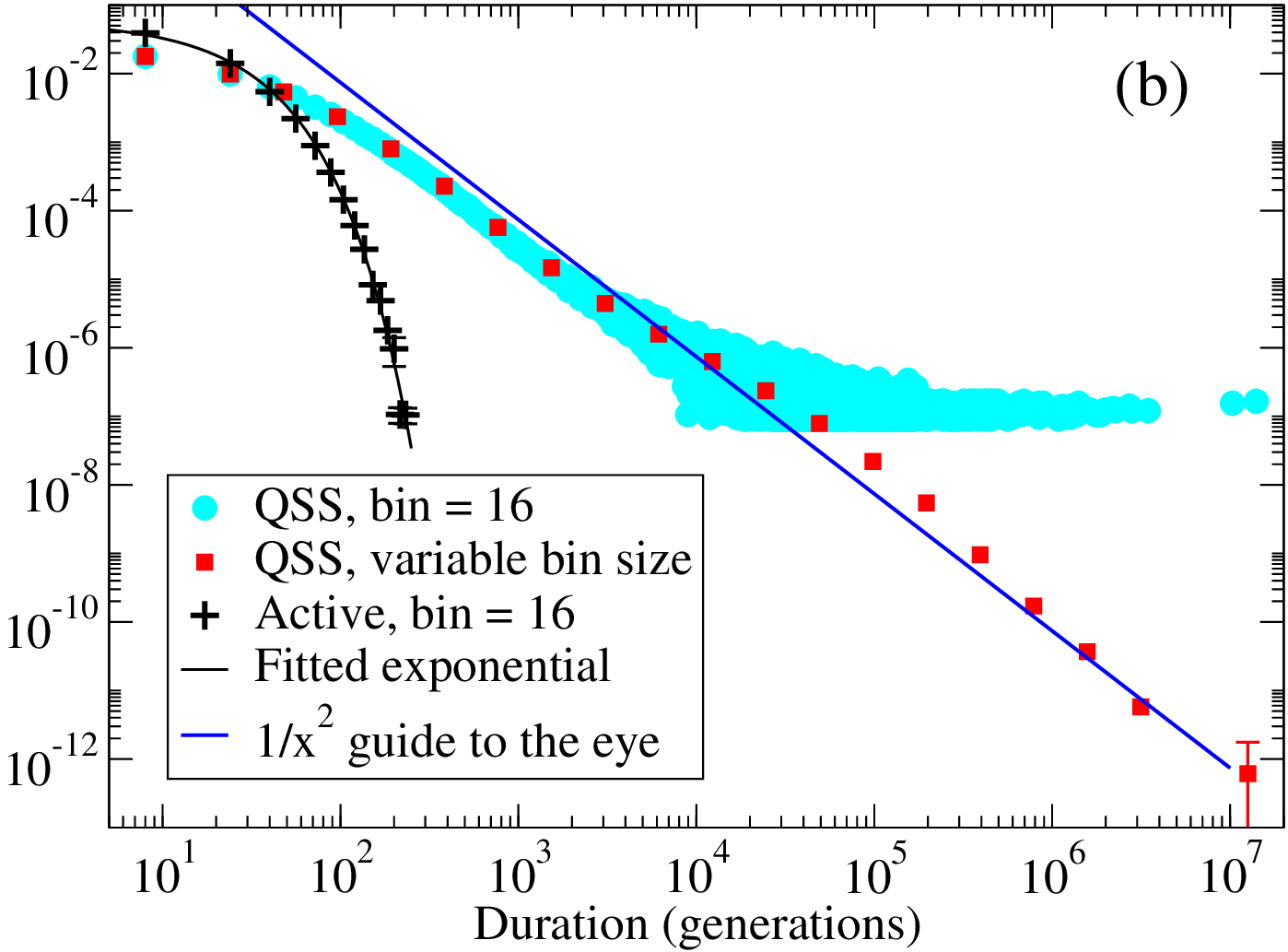} 
\caption[]{
(Color online.) 
(a)
Normalized histogram of entropy changes, averaged over 16 generations 
(thick curve with shading).
Based on 16 independent runs of $2^{25} = 33\,554\,432$ generations each. 
The near-Gaussian
central peak corresponds to the QSS, while the near-exponential wings
correspond to the active periods. Also shown by a thin line is a
histogram based on birth/death 
fluctuations in ten specific QSS communities with zero mutation
rate. The latter is re-normalized so that the maxima of the two
histograms coincide. Based on these histograms, a cutoff of $[S(t) -
S(t-16)]/16 = 0.015$ was used to distinguish between active periods and
QSS. 
(b)
Normalized histograms for the length of active periods ($\bf + $) 
and QSS [solid, light gray circles (cyan online) and solid, dark gray 
squares (red online)]. 
Based on 16 independent runs of $2^{25}$ generations each. 
Two of the histograms ($+$ and circles) use a
constant bin width of 16 generations. In order to capture the information
for large durations, the data for the QSS were also analyzed
with exponentially increasing bin size (squares). 
Error bars showing standard error based on the spread between the
individual runs are shown only where they are larger than the symbol size. 
The black curve through the points for the active periods is a
least-squares fit of an exponential distribution to the data. 
The straight line is a guide to the eye, corresponding to $1/x^2$ behavior 
for the QSS data. 
See Sec.~\protect\ref{sec:Life} for details. 
}
\label{fig:periods}
\end{figure}

The suggestion of self-similarity that emerges from 
Figs.~\ref{fig:timeser} and~\ref{fig:spec} makes it natural to investigate
the statistics of the durations of active periods and QSS over a wider 
range of time scales than that represented by the ten QSS included 
in Table~\ref{table1}. From 
Fig.~\ref{fig:timeser} we see that a reasonable method to make
this distinction is to observe the magnitude of the entropy changes, 
$|d S(t) / d t|$, and to consider the system to be in the active state if
this quantity is above some suitably chosen cutoff. 
This cutoff can best
be determined from a histogram of $d S / d t$, such as shown in 
Fig.~\ref{fig:periods}(a). 
%The histogram is based on 16 independent runs of $2^{25}$ generations, sampled
%every 16 generations. 
It shows that the probability
density of the entropy derivative consists of two additive
parts: a near-Gaussian one corresponding to population fluctuations caused
mostly by the birth/death process in the
QSS, and a second one corresponding to large changes during the active 
periods. From this figure we chose the cutoff as 0.015, which was used to
classify every sampled time point as
either active or quiet. Normalized histograms for the 
durations of the active and quiet periods
%, averaged over the same 16 long simulations, 
are shown in 
Fig.~\ref{fig:periods}(b). About 97.4\% of the time is spent in QSS. 
The active periods are seen to be relatively short, and their probability 
density is fitted well by an exponential distribution \cite{NOTEa}. 
On the other hand, the lifetimes of the QSS show a very broad distribution
\cite{CHRI02} -- possibly a power law with an exponent near $-$2 for
durations longer than about 200 generations. We note that there is some 
evidence of power laws with exponents near $-$2 in the
distributions of several quantities extracted from the fossil record, such
as the life spans of genera or families, and the sizes of extinctions 
\cite{SOLE96,NEWM96,SHIM02,SHIM03}. 
However, the fossil data are sparse and extend over no more than one or
two decades in time, and they can be fitted almost as well 
by exponential distributions \cite{SOLE96}. 
Although our data extend over a much wider range of time scales than the
paleontological data, indisputable evidence of a power law remains to be
established. Though a few QSS of the order of one million generations will
certainly appear in every run of $2^{25}$ generations, 
more definite conclusions about the
statistics of such long QSS must await simulations an order of 
magnitude longer. Another
feature in Fig.~\ref{fig:periods}(b) is that the distribution of
QSS durations changes to a smaller slope below about 200 generations. 
A similar effect has been
observed in the fossil record of lifetimes of families
\cite{SHIM02,SHIM03}. 

\subsection{Power spectral densities}
\label{sec:PSD}

\begin{figure}[ht] 
\includegraphics[angle=0,width=.47\textwidth]{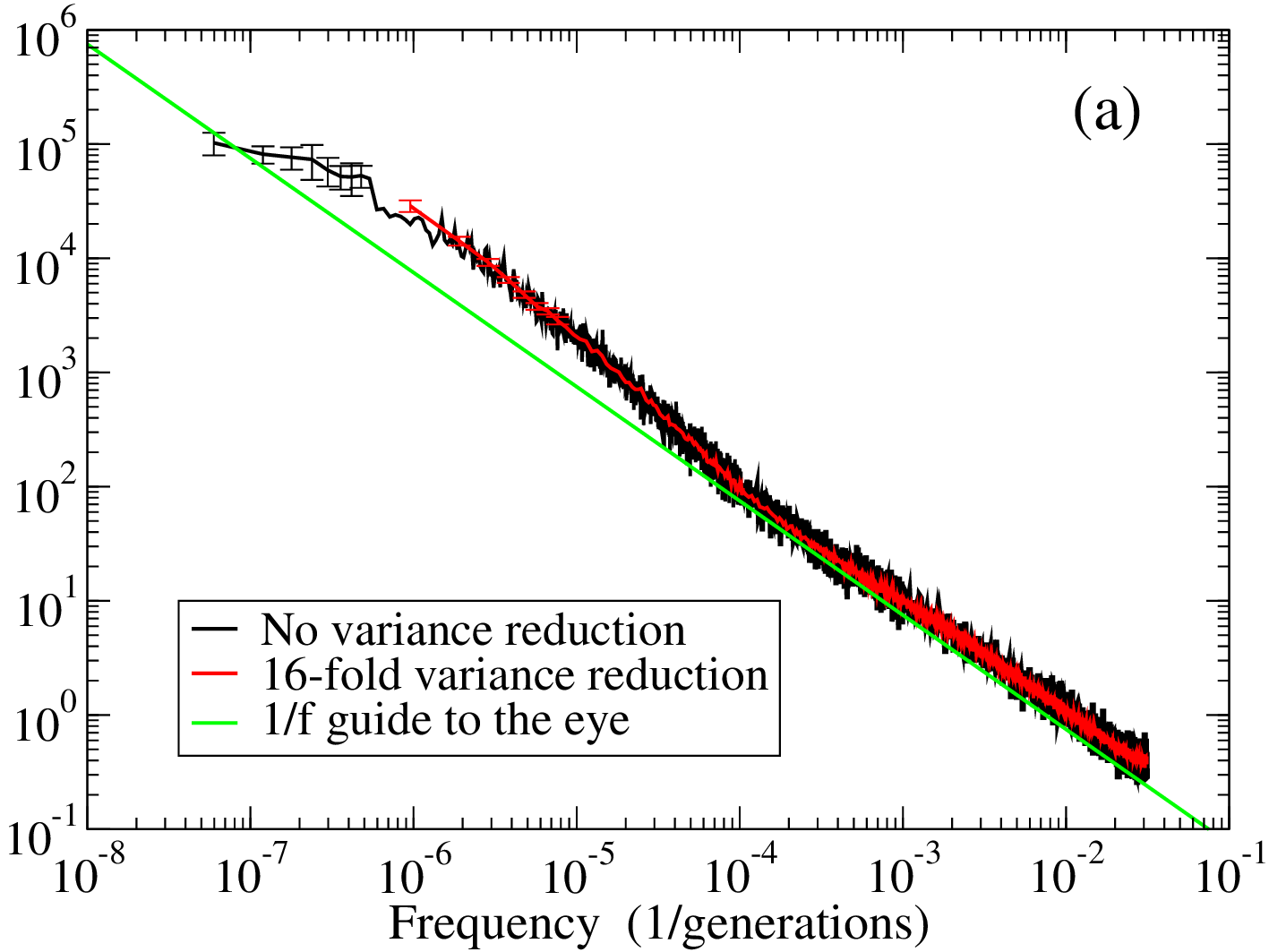} 
\hspace{0.5truecm}
\includegraphics[angle=0,width=.47\textwidth]{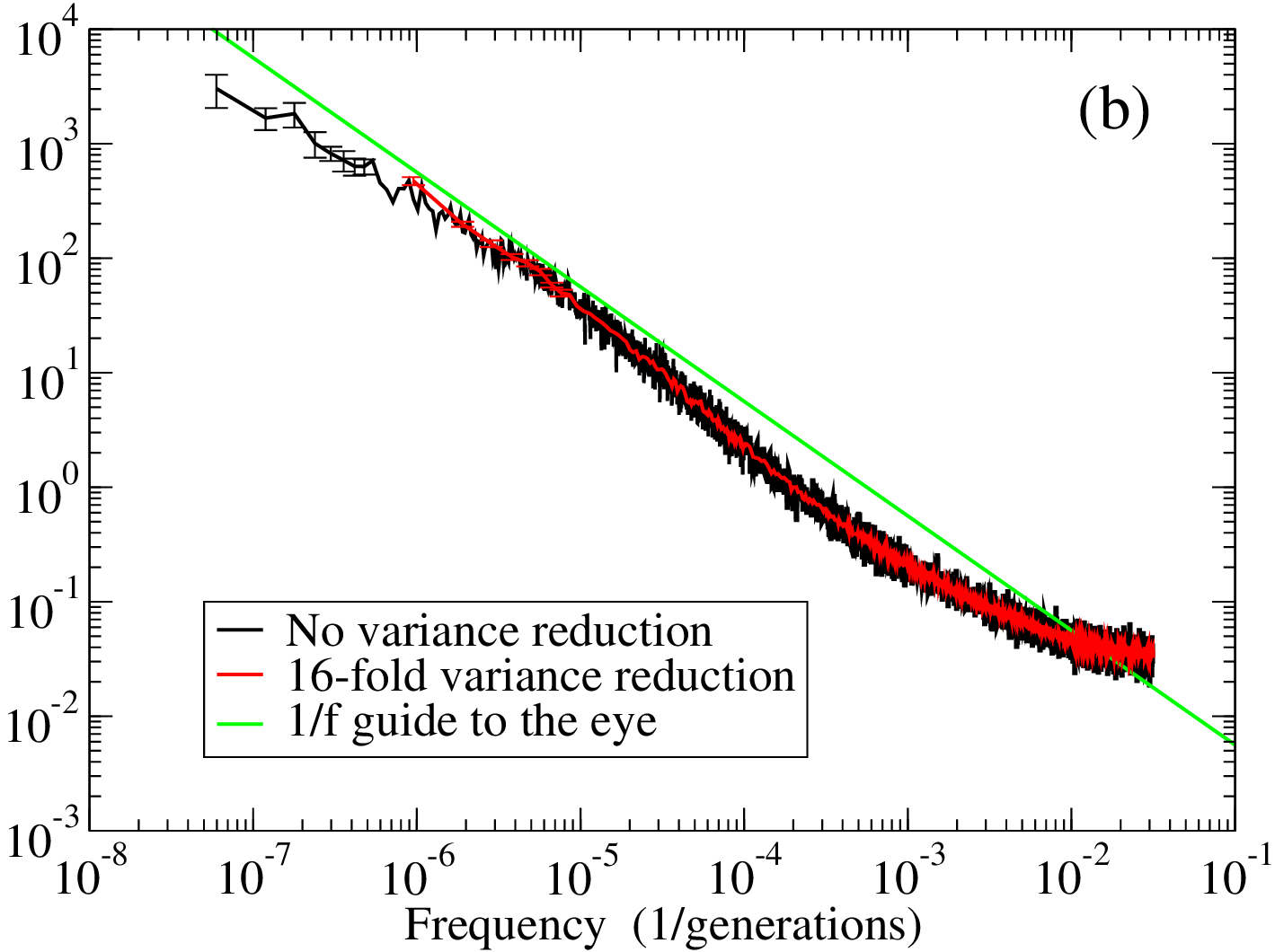} 
\caption[]{
(Color online.) 
Power spectral densities (PSD) \protect\cite{PSD} for simulations of length 
$2^{25}$ generations, sampled every 16 generations and 
averaged over 16 independent runs. 
Black curve: no variance reduction. 
Gray curve (red online): 16-fold variance reduction. 
Error bars shown for the eight lowest frequencies in each curve are standard 
errors, based on the spread between the individual runs. 
The straight line is a guide to the eye, corresponding to $1/f$ behavior.
(a) 
Shannon-Wiener diversity index $D(t) = \exp[S(\{n_I(t)\})]$.
(b)
Normalized total population $N_{\rm tot}(t)/[N_0 \ln (F-1)]$.
See discussion in Sec.~\protect\ref{sec:PSD}. 
}
\label{fig:PSD}
\end{figure}

{}From the discussion in Sec.~\ref{sec:Life}
it is clear that by using a cutoff for the
intensity of some variable which is large in the QSS and small in the
active periods (but in both cases with fluctuations of unknown
distribution), it is impossible to classify the periods unambiguously. For
example, by increasing the cutoff one can make appear as a single QSS
period what previously appeared as two successive, shorter QSS periods 
separated by a short active period. Since the probability of encountering an
extremely large fluctuation in a QSS increases proportionally to the length
of the QSS, this effect will affect the longest
periods most severely. The problem with determining a suitable cutoff can
be avoided by, instead of concentrating on period statistics, 
calculating the power spectral density (PSD) \cite{PSD,PRES92} 
of variables such as the Shannon-Wiener diversity index
or the total population. PSDs for these two quantities are shown in
Fig.~\ref{fig:PSD}. 
%The data are averaged over 16 independent
%runs of $2^{25}$~generations each. 
The PSD for the diversity $D(t)$, shown in Fig.~\ref{fig:PSD}(a),
appears inversely proportional to the frequency $f$ 
($1/f$ noise or flicker noise \cite{MILO02})
for $f > 10^{-4}$, goes through a crossover regime of about one decade
in $f$ where it is $\sim f^{- \alpha}$ with $\alpha > 1$, and then
appears to return to $\alpha \approx 1$ for $10^{-6} < f < 10^{-5}$. For $f <
10^{-6}$ our data are insufficient to determine the PSD unambiguously,
and much longer simulations would be necessary \cite{NOTEb}. 
%XXX test wavelet method! XXX 
In the PSD for $N_{\rm tot}(t)$, shown in Fig.~\ref{fig:PSD}(b), the
substantial population fluctuations due to the birth/death process
during the QSS periods produce a
large noise background which interferes with the interpretation at high
frequencies. However, for $f < 10^{-5}$ the behavior is also 
consistent with $1/f$ noise. 

On time scales much longer than the mean duration of an active period, the
time series for the diversity can be approximated by constants
during the QSS, interrupted by delta-function like spikes corresponding
to the active periods (see Fig.~\ref{fig:timeser}). In this limit, 
the relation between a very wide distribution of the QSS durations $\tau$, 
described by a long-time power-law dependence of the probability density 
$p(\tau) \sim \tau^{- \beta}$, and the 
low-frequency behavior of the PSD is known analytically \cite{PROC83}: 
\begin{equation}
P(f) 
\sim 
\left\{
\begin{array}{lll}
f^{-(\beta - 1)} 
& \mbox{for} & \beta < 2 \nonumber\\
1/\left( f | \ln f |^2 \right)  
& \mbox{for} & \beta =2 \nonumber\\ 
f^{-(\beta -1) (3 - \beta)}  
& \mbox{for} & 2 < \beta < 3 \nonumber\\ 
| \ln f |  
& \mbox{for} & \beta = 3 \nonumber\\ 
{\rm const.}   
& \mbox{for} & \beta > 3  
\end{array}
\right.
\label{eq:PROC}
\end{equation}
Thus, the approximate $1/f$ behavior of the PSDs in Fig.~\ref{fig:PSD}
is consistent with
the approximate $1/\tau^2$ behavior of the probability density for
the QSS durations, shown in Fig.~\ref{fig:periods}(b). 

As well as power-law distributions, PSDs that go as $1/f^\alpha$ with $\alpha
\approx 1$ have been extracted from the fossil record \cite{SOLE97}.
However, such observed PSDs extend only over one to two decades in
frequency (corresponding to power-law probability densities
extending only over one or two decades in time), and more recent
work indicates that the $1/f$ spectra obtained in Ref.~\cite{SOLE97} may be
artifacts of the analysis method \cite{NEWM99}. Although $1/f$ noise,
at least in some frequency interval, 
is a property of our model and is possibly also 
seen in other models of macroevolution \cite{BAK93} 
(but see Refs.~\cite{DAER96,DAVI01} for conflicting opinions on its presence in
the Bak-Sneppen model), whether or not it is really present in the
fossil record remains an open question. 

\subsection{Stationarity and effects of finite genome size}
\label{sec:Stat}

An important issue in evolutionary biology is whether or
not the evolving ecosystem is stationary in a statistical sense. 
In Fig.~\ref{fig:RunAve} we show several diversity-related measures, 
averaged over a moving time window and independent simulation runs. 
These quantities are: 
the species richness ${\cal N}(t)$, 
the number ${\cal N}_2(t)$ of genotypes with $n_I \ge 2$, 
the Shannon-Wiener index $D(t)$, 
the total population $N_{\rm tot}(t)$, 
and the average Hamming distance $\langle H \rangle$ 
between genotypes in the community and its standard deviation $\sigma_H$. 
As seen in the figure, none of these quantities 
show any signs of a long-time trend over 
$2^{25}$ generations. This is consistent with fossil evidence for
constant diversity \cite{MAYN89}. However, it is in disagreement with
simulations of the original version of the tangled-nature model
\cite{HALL02}, in which a slow growth in species richness was
observed. This discrepancy might possibly result from the different
forms of the interaction matrix used in the two studies, but it may also
be due to the relatively short time series used in Ref.~\cite{HALL02}.  

{}For our model, a simple, combinatorial phase-space
argument (containing neither the individual 
population sizes nor the mutation rate) indicates that the most probable number
of major genotypes should indeed be limited. The argument goes as follows. 
A community of $\cal N$ genotypes can be chosen in 
$2^L \choose {\cal N}$ different ways and is influenced by 
${\cal N}({\cal N} -1)/2$ different pairs of $M_{IJ}$ and $M_{JI}$. 
We let the probability that the pair of interactions is suitable to
forming a stable community be $q$ (if the requirement is simply that both 
$M_{IJ}$ and $M_{JI}$ should be positive, 
then $q=1/4$ with our choice of $\bf M$).
Thus, a stable $\cal N$-species community can be formed in approximately
\begin{equation}
\Omega({\cal N}) 
= 
{2^L \choose {\cal N}} 
q^{{\cal N}({\cal N} -1)/2} 
\label{eq:Omega}
\end{equation}
ways, which obeys the recursion relation 
\begin{equation}
\Omega ({{\cal N}+1}) 
= 
\frac{2^L - {\cal N}}{{\cal N} + 1} q^{\cal N} \Omega ({\cal N})
\label{eq:pn}
\end{equation}
with $\Omega(1) = 2^L$. 
This recursion relation can either be used to find the most probable
value ${\cal N}^{\,\dag}$ numerically, or one can easily obtain an estimate
valid for $2^L \gg {\cal N}$: ${\cal N}^{\,\dag} \approx L \ln 2 / \ln (1/q)$. 
The value numerically obtained with $q=1/4$ is ${\cal N}^{\,\dag} = 6$,
but the close correspondence to the results shown in
Fig.~\ref{fig:RunAve} may be fortuitous since the dependence of ${\cal
N}^{\,\dag}$ on $q$ is only logarithmic. 

\begin{figure}[ht] 
\includegraphics[angle=0,width=.47\textwidth]{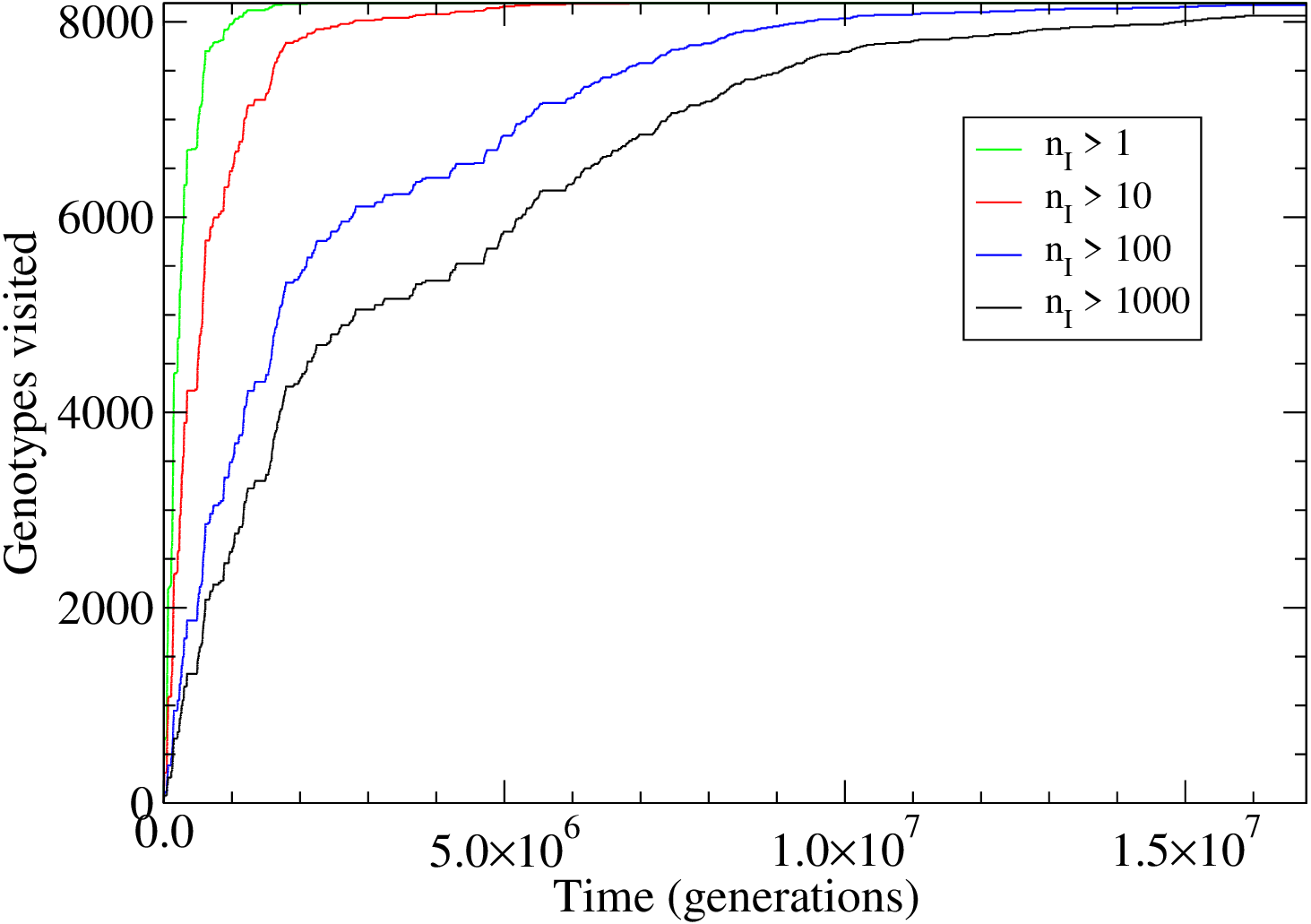} 
\caption[]{
(Color online.) 
The number of different 
genotypes visited at different population levels, shown vs
time for the first half of a single run of $2^{25}$~generations. 
From above to below, $n_I \ge 2$, $n_I \ge 11$, $n_I \ge 101$, 
and $n_I \ge 1001$. Horizontal plateaus correspond to QSS. 
See discussion in Sec.~\protect\ref{sec:Stat}. 
}
\label{fig:SpecVis}
\end{figure}

Another question of interest is whether the limited size of the
genome leads to ``revisiting'' of genotypes and communities. The answer
is affirmative and indicates the importance of further studies of the
effects of the genome size on the long-time dynamical behavior. 
For genotypes the
question of revisits is easy to answer, and the results for a single
run of $2^{25}$ generations are shown in Fig.~\ref{fig:SpecVis}. 
As we see, in less than $3\times 10^6$ generations, at least two individuals
of every genotype have appeared simultaneously (curve labeled $n_I > 1$).
By the end of the run, almost every genotype has enjoyed being a major species
with $n_I \ge 1001$ at least once.
Plots on a log versus linear scale (not shown) indicate that the curves 
are reasonably well
approximated by exponential approaches to ${\cal N}_{\rm max}$, 
indicating that genotypes appear to be nearly
randomly visited and revisited at a constant rate dependent on $n_I$. 

For communities, the question is more difficult. If a community is simply
defined as an internally stable ${\cal N}$-species community, then an
estimate for their total number can be obtained from $\Omega ({\cal N})$ of
Eq.~(\ref{eq:Omega}) as described in Appendix~\ref{sec:AC}. The result with 
the parameter values used in this paper is of the order of $10^{12}$.
However, unstable communities are also briefly visited, especially during
active periods, and a more inclusive way of counting communities would be a
coarse-graining procedure based on the overlap function defined in Eq.~(\ref
{eq:over}). The nonzero populations are recorded at $t=0$, and the overlap
with this initial community is monitored until at some $t^{\prime }>0$ it
falls below a suitably chosen cutoff ${\cal O}_{{\rm cut}}$. The set 
$\{n_I(t^{\prime })\}$ is then recorded as a new community with $t^{\prime }$
as its starting time. The process is repeated, now comparing with the newly
recorded community. This procedure creates a list of communities and
their starting times. To address the issue of revisits, 
we next scan this list to extract those communities that
represent revisits to a previously visited community, or
``prototype community.'' We compare each community 
$\{n_I(t)\}$ in the original list with the previously found communities 
$\{n_I(\tilde{t})\}$, sequentially in order of increasing $\tilde{t} < t$.  
If none of the overlaps is greater than ${\cal O}_{{\rm cut}}$, 
$\{n_I(t)\}$ is added to the list of prototypes 
${\cal M}_{\rm p}(t_{\rm p})$ 
with $t_{\rm p} = t$. If an overlap
greater than ${\cal O}_{{\rm cut}}$ is found at some $\tilde{t}$, we stop the  
comparisons and add this community to the list of revisits 
${\cal M}_{\rm r}(t_{\rm r})$. The starting time of this revisited
community, $t_{\rm r} = t$, 
is associated in the list ${\cal M}_{\rm r}(t_{\rm r})$
with the unique $t_{\rm p} = \tilde{t}$, 
the starting time of the associated prototype. 
As the system evolves, the number of items in each list
increases monotonically. Clearly, the cutoffs should be
sufficiently small that communities that differ only by fluctuations inside
a QSS are not considered different, but sufficiently large to avoid counting
significantly different communities as identical. From inspection of the overlap
fluctuations in some of the QSS included in Table~\ref{table1}, we found 
that ${\cal O}_{{\rm cut}}$ in the range of 0.90 to 0.95 is optimal. 
Estimates of
the total numbers of communities (stable and unstable) that can in principle
be distinguished by this method are obtained in Appendix~\ref{sec:AC}.
Depending on details of the assumptions, the estimates vary between $10^{21}$
and $10^{28}$ for these cutoffs.

\begin{figure}[ht] 
\includegraphics[angle=0,width=.47\textwidth]{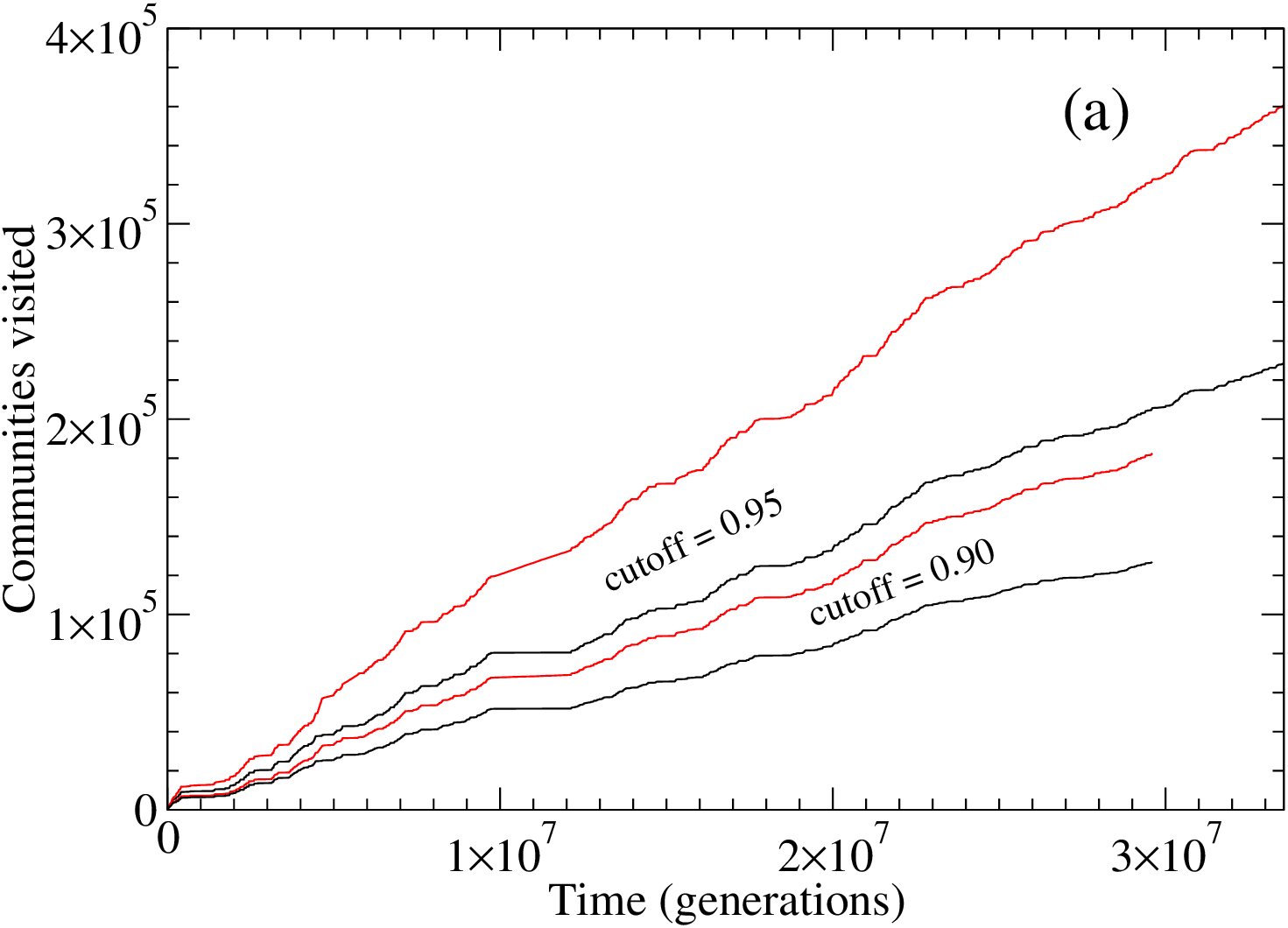} 
\includegraphics[angle=0,width=.47\textwidth]{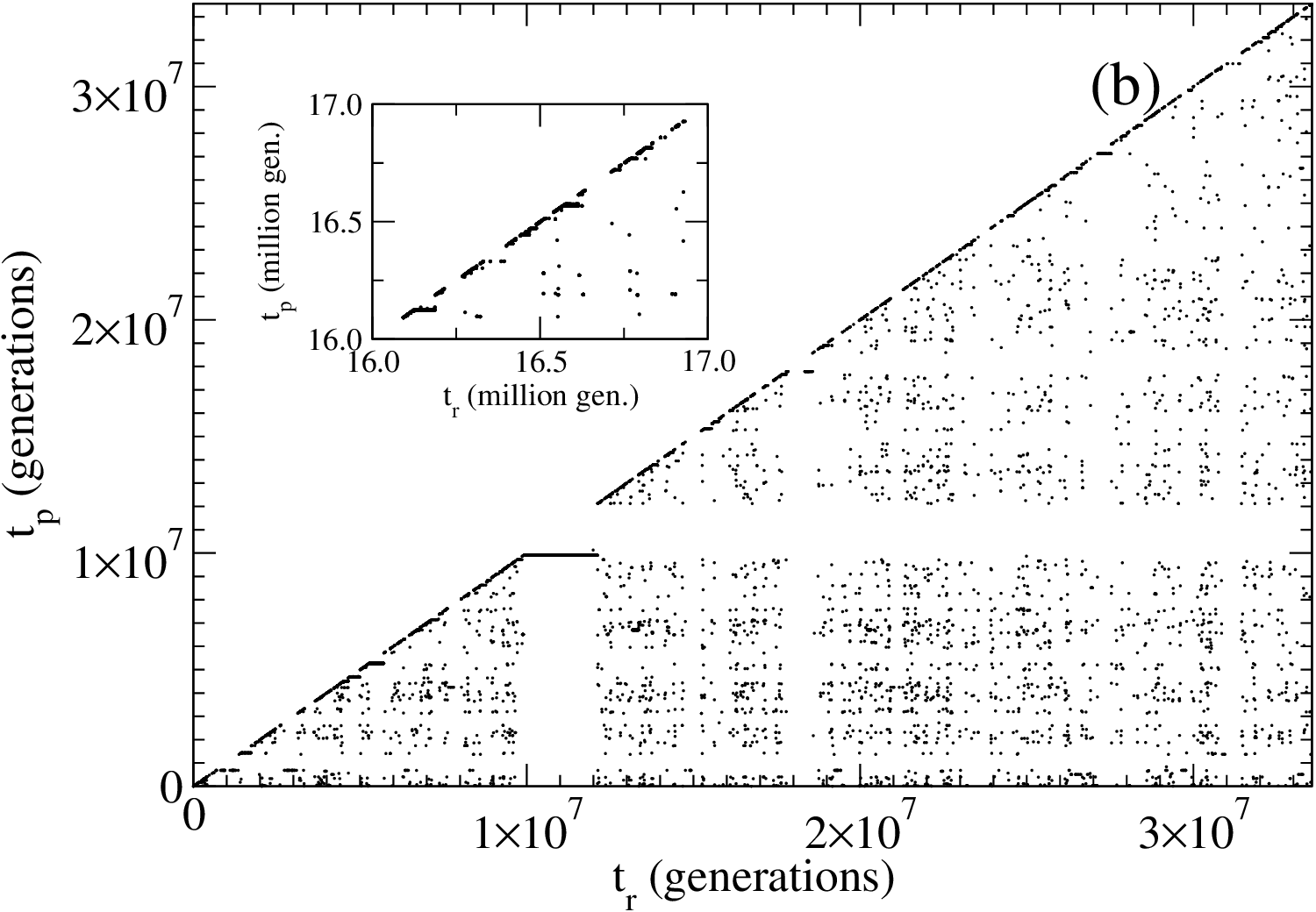} \\
\vspace*{1.0truecm}
\includegraphics[angle=0,width=.47\textwidth]{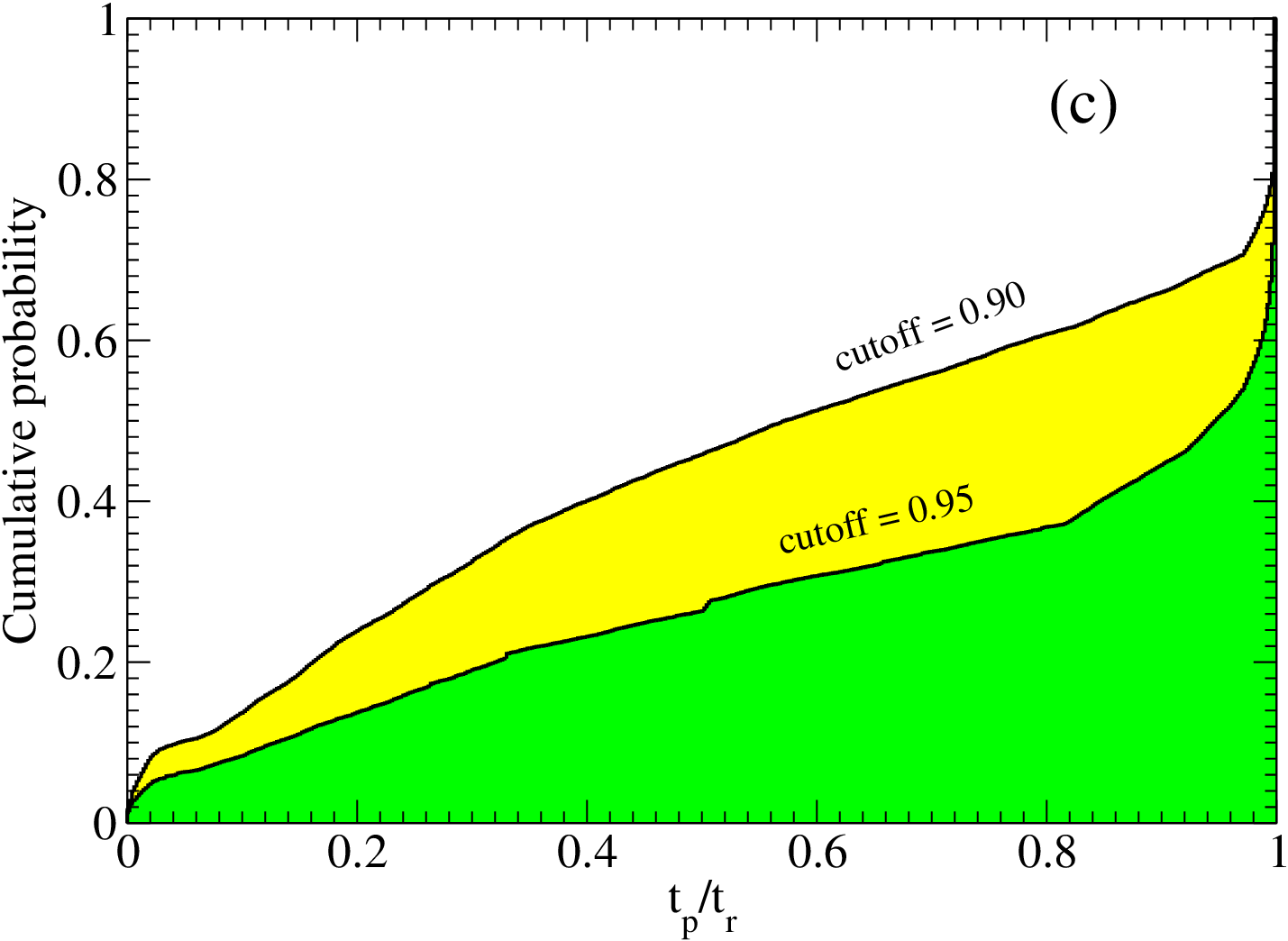} 
\caption[]{
(Color online.) 
(a)
Number of communities visited, shown vs
time for a single run of $2^{25}$ generations. A new community is
counted whenever the overlap function falls below  
${\cal O}_{\rm cut}$. The upper
pair of curves corresponds to ${\cal O}_{\rm cut} = 0.95$, 
and the lower pair to 0.90. 
The top curve in each pair (thin, gray curve, red online)  
counts both prototype communities and revisits, 
while the bottom curve (heavy, black curve) excludes revisits. 
(b)
{}For ${\cal O}_{\rm cut} = 0.95$ only, the abscissa 
$t_{\rm r}$ gives the starting time of each revisit, while 
the ordinate $t_{\rm p}$ gives the 
starting time of the associated prototype community.
Inset: 
detail for $t_{\rm p} \alt t_{\rm r}$ over $10^6$ generations. 
(c)
Lower curve: 
cumulative histogram for $t_{\rm p}/t_{\rm r}$ from part (b). 
About 60~\% of the
revisits are to other recently visited communities, while
the rest are approximately uniformly distributed over all previously
visited communities. 
Upper curve:
corresponding result for ${\cal O}_{\rm cut} = 0.90$.
See discussion in Sec.~\protect\ref{sec:Stat}. 
}
\label{fig:CommVis}
\end{figure}

The results of the procedure described above with ${\cal O}_{{\rm cut}}=0.90$
and $0.95$ are shown in Fig.~\ref{fig:CommVis} for a single run of $2^{25}$
generations. The upper pair of curves in Fig.~\ref{fig:CommVis}(a)
corresponds to ${\cal O}_{{\rm cut}}=0.95$, and the lower pair to 
${\cal O}_{ {\rm cut}}=0.90$. 
In each pair, the upper curve shows the total number of
communities in the lists ${\cal M}_{\rm p}\left( t_{\rm p} \right)$ and 
${\cal M}_{\rm r}\left( t_{\rm r} \right) $, while
the lower curve shows just the number of prototypes 
in ${\cal M}_{\rm p}\left( t_{\rm p} \right) $. 
For ${\cal O}_{{\rm cut }}=0.95$, 
about 40\% of the communities are seen to be revisits, 
%(i.e., ${\cal M}_p/{\cal M}_r\sim 1.5$), 
while the proportion is about 30\%
%(${\cal M }_p/{\cal M}_r\sim 2 $) 
for ${\cal O}_{{\rm cut}}=0.90$. In both cases, 
the curve for the number of prototypes
remains approximately linear, indicating that
the supply of previously unvisited communities is nowhere near to being
exhausted, even for such a long run. In view of the enormous numbers of
available communities estimated above, this result is reasonable. QSS appear
as plateaus in the curves showing the 
numbers of prototypes. Figure~\ref{fig:CommVis}(b) 
provides a different perspective (for ${\cal O}_{{\rm cut}}=0.95$ only).
Each revisit is represented by a point with $t_{\rm r}$ as 
abscissa and $t_{\rm p}$ as ordinate. 
Thus, every point lies strictly below the
diagonal. Long-lived QSS appear as large gaps and horizontal
segments (e.g., the one near $10^7$ generations). 
The inset shows a detail of $10^6$
generations near the diagonal around $1.65\times 10^7$. As we see, there are
many points{\em \ just below }the diagonal, 
representing the fluctuations around the
(many short-lived) QSS. By contrast, points far below the diagonal
represent ``throwbacks'' to the vicinity of earlier prototype communities.
Note that the density of points is much higher just below the diagonal,
implying that a large portion of the revisited communities are ``fluctuation
related.'' To highlight these differences, we show the cumulative
probability of the ratio $t_{\rm p}/t_{\rm r}$ 
in Fig.~\ref{fig:CommVis}(c), in which the lower
curve corresponds to the data in Fig.~\ref{fig:CommVis}(b). For the case of
${\cal O}{{\rm cut}}=0.95$ (lower curve), we see that about 60\% of the
revisits can be regarded as ``fluctuation related'' (with 
$t_{\rm p} \alt t_{\rm r}$), 
while the rest are ``throwbacks.'' Roughly, the latter
component appears to be distributed uniformly over all earlier times.  The
upper curve shows the corresponding result for ${\cal O}_{{\rm cut}}=0.90$.
Corresponding to using a more coarse-grained covering of 
state space, it naturally
displays a larger proportion of throwbacks. Not surprisingly, there is no
sharp distinction between these two components of the revisited communities
since even a rough partition depends on the details of coarse graining.
Nevertheless, we can conclude that the dynamics produces a steady stream of
essentially new communities drawn from the vast supply of possibilities.

\section{Summary and Conclusions}
\label{sec:Concl}

In this work we have studied, by linear stability analysis and large-scale
Monte Carlo simulations, a simplified version of the tangled-nature model of
biological coevolution, recently introduced by 
Hall, Christensen, and collaborators \cite{HALL02,CHRI02,COLL03}. 
Selection is provided by interspecies interactions through the 
reproduction probability $P_I$ [Eq.~(\ref{eq:P})], 
which corresponds to a nonlinear population-dynamics model of the 
community ecology, while the genetic 
variability necessary for evolution is provided by a low rate of mutations 
that act on individual organisms during reproduction. 

At the low mutation rate studied here, the model provides an intermittent,
statistically self-similar behavior, characterized by periods
of relative calm, interrupted by bursts of rapid turnover in genotype space.
During the quiet periods, or quasi-steady states (QSS), the population
consists of a community of a relatively small number of 
mutualistically interacting genotypes. The populations of the 
individual genotypes, $n_I(t)$, fluctuate near a stable fixed
point of a deterministic mean-field, mutation-free version of the 
model [Eq.~(\ref{eq:mu=0})]. During the active
periods, the system moves through genotype space at a rate that is rapid on 
a macroscopic (``geological'') time scale, although of course finite on a 
microscopic (``ecological'') scale. These periods of rapid change are 
characterized by large
fluctuations in the diversity and an overall reduction of the 
total population. In long simulation runs of
$2^{25}$~generations, the ecosystem spends about 97.4\% of the time in QSS. 
The time series produced by the model are statistically stationary, and there
is no evidence that any particular quantity is being optimized as the system
moves through genotype space. In that sense, the dynamical behavior
can probably best be described as a neutral drift \cite{DROS01}. 
Overall, the dynamical behavior 
of this model resembles closely the punctuated-equilibria
mode of evolution, proposed by Gould and Eldredge \cite{GOUL77,GOUL93,NEWM85}.

Investigation of duration statistics for the quiet periods shows a very
wide distribution with a power-law like long-time behavior characterized by 
an exponent near $-2$. Consistent with this result, the 
power spectral density of the diversity shows $1/f$ noise. 
While there are claims that similar statistics 
characterize the fossil record \cite{SOLE97,SOLE96,NEWM96,SHIM02,SHIM03},
this is still a contested issue \cite{SOLE96,NEWM99}. At best, observations of
power-law distributions and $1/f$ noise in the fossil record extend over no
more than one or two decades in time or frequency, and it must remain an open
question whether this is the optimal interpretation of the scarce data
available. 

Due to the absence of sexual reproduction, our model can at best be
applied to the evolution of asexual, haploid organisms such as bacteria. 
It should also be noted that no specific, biologically relevant information
has been included in the interaction matrix $\bf M$. 
In particular, this fact may be responsible for
our QSS being strongly dominated by mutualistic relationships. 
The absence of biologically motivated detail in $\bf M$ is both a strength and
a weakness of the model. 
Its strength lies in reinforcing the notion of universality in
macroevolution models, e.g., power law behaviors and $1/f$ noise. 
By the same token, its weakness lies in the lack of biological detail, to
the point that comparison with specific observational or experimental data
is difficult. Clearly, the detailed effects of interspecies interactions
on the macroevolutionary behavior in models similar to the one studied here
represents an important field of future research. Examples include
the importance of the connectivity of the interaction matrix, 
correlated interspecies interactions \cite{SEVI04}, and
interaction structures corresponding to food webs with distinct trophic levels. 

Despite all these caveats, we find it encouraging that such
a simple model of coevolution with individual-based dynamics
can produce punctuated equilibria, power-law distributions, and $1/f$ noise
consistent with current theories of biological macroevolution. We believe
future research should proceed in the direction
pointed out by this and similar models. This entails 
combining stochastic models from community ecology with models of 
mutations and sexual reproduction at the level of individual organisms, 
and investigating the consequences of more
biologically realistic interspecies interactions.

\section*{Acknowledgments}
\label{sec:ACK}

We are grateful for many helpful comments on the
manuscript by J.~A.\ Travis,
and we acknowledge useful conversations and correspondence with 
I.~Abou Hamad,
G.~Brown,
T.~F.\ Hansen, 
N.~Ito,
H.~J.\ Jensen,
A.~Kolakowska,
S.~J.\ Mitchell,
M.~A.\ Novotny,
H.~L.\ Richards,
B.~Schmittmann,
V.~Sevim,
U.~T{\"a}uber, 
and 
J.~C.\ Wilgenbusch.

P.~A.~R.\ appreciates the hospitality of the 
Department of Physics, Virginia Polytechnic Institute and State University, 
and the Department of Physics and Astronomy and ERC
Center for Computational Sciences, Mississippi State University.

The research was supported in part by National Science Foundation Grant 
Nos.~DMR-9981815, DMR-0088451, DMR-0120310, and DMR-0240078,
and by Florida State 
University through the Center for Materials Research and Technology and 
the School of Computational Science and Information Technology.

\appendix

\section{Master equation}
\label{sec:AB}

A complete description of the stochastic process can be given in terms of a
master equation, which specifies the evolution of 
${\cal P}\left( \vec{n},t\right)$, 
the probability that the system is found with composition $\vec{n}$ at time 
$t$. Here, 
$\vec{n} \equiv \left\{ n_1,n_2,...,n_{{\cal N}_{\rm max}}\right\}$, and
${\cal N}_{\rm max} = 2^L$ 
is the number of individuals of species $I$. In our case, 
$L=13$ and ${\cal N}_{\rm max} = 8192$. 
Similar to the main text, we define 
$N_{\rm tot}\equiv \sum_In_I $
and let $N_0$ be the carrying capacity. 

We write the probability for an
individual of species $I$ to survive to reproduce as 
\begin{equation}
P_I\left( \vec{n}\right) =\left\{ 1+\exp \left[ \frac {N_{\rm tot}}{N_0} 
- \sum_JM_{IJ} \frac{n_J}{N_{\rm tot}}\right] \right\} ^{-1} .
\label{eq:B1}
\end{equation}
The main difference between this expression and Eq.~(\ref{eq:P}) lies in the
interpretation. Here, $\vec{n}$ is a ``coordinate variable'' in the 
${\cal N}_{\rm max}$-dimensional space, in contrast to $n_I\left( t\right) $ 
being just a point in this space. 

To proceed, we define the symbol 
$ {n_I \atopwithdelims[] m_I}$
as the rate for survival (from $n_I$ individuals to $m_I$ ``mothers'').
Since each individual is given a chance to survive according to $P_I$, we
have 
\begin{equation}
{n_I \atopwithdelims[] m_I} 
=\frac{n_I!}{m_I!\left( n_I-m_I\right) !}\left( P_I\right) ^{m_I}\left(
1-P_I\right) ^{n_I-m_I} , 
\label{eq:B2}
\end{equation}
which has the form of a binomial distribution. 
Next, each mother gives rise to $F$ offspring. However, due to mutations,
not every offspring is of the same genotype as the mother. Although it is
possible to have mutants with a genotype differing from the mother by more
than one bit, we restrict ourselves here to a simpler version, namely
mutant genotypes that can differ only by one bit. 
Since our simulations typically
involve $\mu \sim 10^{-3}$, this restriction should not lead to serious
difficulties. Given that only one bit may be flipped, there are $L+1$
possible varieties of offspring for each maternal genotype. 
We introduce the notation 
\begin{eqnarray*}
&&b_{J,0}\quad \text{for the number of 
offspring from mother }J\text{ with no mutations}
\\
&&b_{J,k}\quad \text{for the number of 
offspring from mother }J\text{ with the }k\text{th bit flipped}
\end{eqnarray*}
We now define the multinomial-like symbol 
\begin{equation}
{
{Fm_J \atopwithdelims[] b_{J,0},b_{J,1},...,b_{J,L}}
} 
=\frac{\left( Fm_J\right) !}{\left( b_{J,0}\right) !}\left( 1-\mu \right)
^{b_{J,0}}\prod_{k=1}^L\frac 1{\left( b_{J,k}\right) !}\left( \frac \mu L
\right) ^{b_{J,k}} ,
\label{eq:B3}
\end{equation}
which is the probability that the $Fm_J$ offspring are distributed
into the specific set $\left\{ b_{J,0},b_{J,1},...,b_{J,L}\right\} $. The
last ingredient needed is the connection matrix 
\begin{equation}
\Delta _G^{J,k}=\left\{ 
\begin{array}{cc}
1 & \text{if genotype }G\text{ is }J\text{ with the }k\text{th bit flipped}
\\ 
0 & \text{otherwise}
\end{array}
\right. 
\label{eq:B4}
\end{equation}
so that the number of offspring 
born {\em into} species $G$ due to mutations is 
\begin{equation}
\tilde{b}_G\equiv \sum_{J,k}\Delta _G^{J,k}b_{J,k} \;. 
\label{eq:B5}
\end{equation}
The final result is 
\begin{equation}
{\cal P}\left( \vec{n}^{\, \prime },t+1\right) =\sum_{\vec{n},\vec{m},\left\{
b\right\} }\prod_G\delta \left( n_G^{\prime }-b_{G,0}-\tilde{b}_G\right) 
\prod_J{ {Fm_J \atopwithdelims[] b_{J,0},b_{J,1},...,b_{J,L}} } 
\prod_I{ {n_I \atopwithdelims[] m_I} }
{\cal P}\left( \vec{n},t\right)
\;.
\label{eq:B6}
\end{equation}

In principle, once the dynamics of $\cal P$ is found, the time dependence of
various quantities (e.g., averages and correlations) can be computed. 
For instance,  
\begin{equation}
\left\langle n_I\right\rangle _t\equiv \sum_{\vec{n}}n_I{\cal P}\left( \vec{n%
},t\right) 
\label{eq:B7}
\end{equation}
is the average number of individuals of species $I$ at time $t$.
From this very detailed description, we see that the evolution of
various quantities can be derived. However, these evolution equations are
extremely complex in general and progress is usually possible only by making
a ``mean-field'' approximation, in which all correlations are neglected.
Thus, averages of products of $\vec{n}$ are replaced by the products of the
averages. As a result, a non-linear evolution equation for $\left\langle 
\vec{n}\right\rangle _t$ results. In a subsequent paper 
\cite{ZIA04}, we shall show that
Eq.~(\ref{eq:EOM}) is precisely the result of such an approach (with $
\left\langle \vec{n}\right\rangle _t$ denoted by $\{ n_I\left( t\right) \}$).
We shall also demonstrate that, for a QSS, ${\cal P}\left( \vec{n}
,t\rightarrow \infty \right) $ is well approximated by a Gaussian (for $\mu
,1/N_0\ll 1$), the center of which is just the fixed point given by 
Eq.~(\ref{eq:fix5}). 
The width of this Gaussian is governed in part by the 
community matrix $\tilde{{\bf \Lambda }}$ of Eq.~(\ref{eq:A2}).
In addition to this (systematic,
``drift'') part, there is a part governing the noise correlations. Between
the two, we can compute distributions of step sizes ($\vec{n}^{\, \prime }-
\vec{n}$). For a long-lived QSS, it is easy to compile data so that 
quantitative comparisons with these predictions can be made.

\section{Counting communities}
\label{sec:AC}

In this Appendix we obtain estimates for the numbers of different
communities that could in principle be observed in an infinitely long
simulation. Even our most conservative estimate is vastly larger than
the numbers observed in a simulation run of $2^{25}$ generations in 
Sec.~\ref{sec:Stat}. 

If a community is simply defined as an internally stable $\cal N$-species
community, then an estimate for their total number can be obtained from 
$\Omega({\cal N})$ of Eq.~(\ref{eq:Omega}). Since $\Omega({\cal N})$ is quite
sharply peaked around ${\cal N}^{\,\dag}$, we can estimate that 
$\sum_{\cal N} \Omega({\cal N}) \sim \Omega({\cal N}^{\,\dag}) 
\approx 2^{L {\cal N}^{\,\dag}} q^{{\cal N}^{\,\dag} ({\cal N}^{\,\dag} -1)/2}
\approx \left( 2^{a/2} \right)^{aL^2}$ with $a = \ln 2 / \ln (1/q)$. For
$q=1/4$ and $L=13$, 
this gives $\Omega({\cal N}^{\,\dag}) \approx 5.2 \times 10^{12}$, while 
direct summation of $\Omega( {\cal N} )$ with the same parameters
gives $\sum_{\cal N} \Omega({\cal N}) = 8.5 \times 10^{11}$ with 46\% due to
the  contribution from ${\cal N} = {\cal N}^{\,\dag} = 6$ and convergence by 
${\cal N} = 11$. As it does not include unstable communities, 
this would serve as our most conservative estimate of the total number
of communities that could be visited in an infinitely long simulation. 

\begin{figure}[ht] 
\includegraphics[angle=0,width=.47\textwidth]{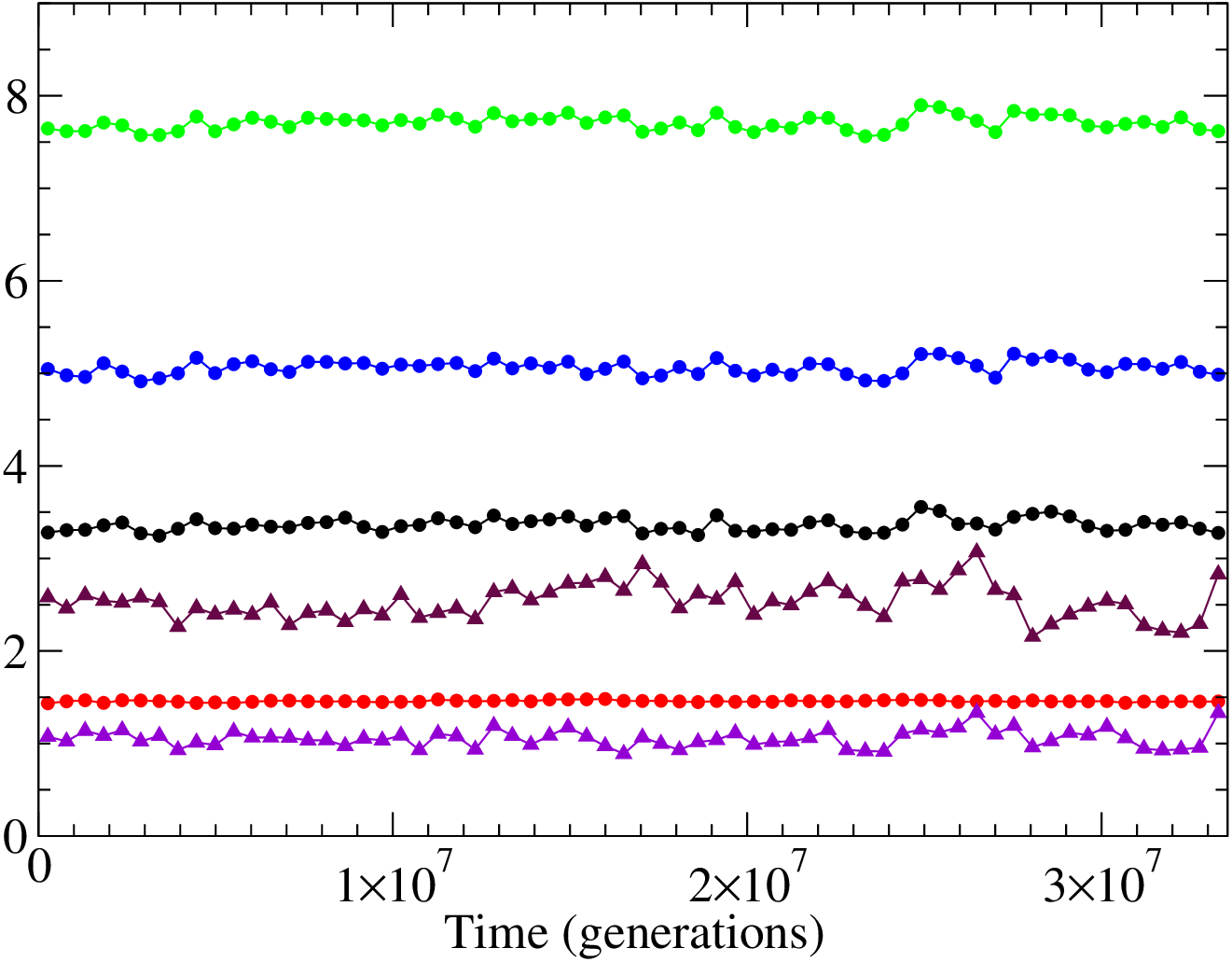} 
\caption[]{
(Color online.) 
Combined time-window and ensemble averages of various 
diversity-related measures. Each data point represents 
an average over a time window of $2^{19} = 524\,288$ generations and 
over 16 (circles) or seven (triangles)
independent runs of $2^{25}$ generations each. 
Shown are, from above to below, the total number of species ${\cal N}(t)$, 
the number ${\cal N}_2(t)$ of species with $n_I \ge 2$,
the Shannon-Wiener index $D(t)$, the average Hamming distance 
$\langle H \rangle$ 
between genotypes in the community, the normalized total population 
$N_{\rm tot}(t)/[N_0 \ln (F-1)]$, and the standard deviation $\sigma_H$ 
of the Hamming distances. These data indicate the absence of any 
systematic long-time trends in the dynamical behavior. 
See discussion in Sec.~\protect\ref{sec:Stat}. 
}
\label{fig:RunAve}
\end{figure}

To estimate the total number of different communities that can be
distinguished by the overlap cutoff method described in
Sec.~\ref{sec:Stat}, we note that 
the overlap threshold ${\cal O}_{{\rm cut}}$ is simply the cosine of the
angle between two vectors in population space, ${\cal O}_{{\rm cut}}=\cos
\theta $. For convenience, we define $\epsilon \equiv 1-{\cal O}_{{\rm cut}}$,
such that $\epsilon $ is small ($0.1$ and $0.05$). Then, $\theta =\arccos
(1-\epsilon )\approx \sqrt{2\epsilon }$.
The fact that the number of populated
species at any one time is near ${\cal N}^\dag$, means that the
direction of the population vector $\{ n_I \}$ to a reasonable
approximation lies within the first
hyper-octant of one of ${\cal N}_{\rm tot} \choose {\cal N}^\dag$
${\cal N}^\dag$-dimensional hyperspheres. 
The total number of communities that can be distinguished with a 
cutoff ${\cal O}_{\rm cut}$
is the ratio of the total ${\cal N}^\dag$-dimensional
solid angle covered by the hyper-octants to that of a hyper-cone
subtended by $\theta$. The solid angle of the hyper-octants is 
\begin{equation}
V_{\rm tot} \sim {{\cal N}_{\rm tot} \choose {\cal N}^\dag}
\frac{S_{{\cal N}^\dag}}{2^{{\cal N}^\dag}} \;,
\label{eq:AC1}
\end{equation}
where $S_d = 2 \pi^{d/2}/\Gamma(d/2)$ is the surface area of
a $d$-dimensional unit sphere. 
The solid angle of the hyper-cone is 
\begin{equation}
C(\epsilon) = S_{{\cal N}^\dag - 1} 
\int_0^\theta (\sin \beta )^{{\cal N}^\dag - 2}
\approx 
S_{{\cal N}^\dag - 1} 
\frac{(2 \epsilon)^{({\cal N}^\dag - 1)/2}}{{\cal N}^\dag - 1}
\label{eq:AC2}
\end{equation}
for small $\epsilon$. 
Using Stirling's approximation for the gamma functions in $S_d$, 
$\Gamma (z) \sim (z/e)^z \sqrt{2 \pi / z}$ and the identity 
$\lim_{n \rightarrow \infty} \left( 1 - \frac{1}{n} \right)^n = e^{-1}$, 
we estimate the total number of different communities to be
\begin{equation}
M( \epsilon ) \sim {{\cal N}_{\rm max} \choose {\cal N}^\dag} 
\frac{\sqrt{2 \pi {\cal N}^\dag \epsilon}}{(2 \epsilon)^{{\cal N}^\dag / 2}}
\;.
\label{eq:AC3}
\end{equation}
This estimate is reasonable only as long as $\epsilon$ is large enough
to exclude minor fluctuations within QSS. 
Observation of overlap fluctuations in some of the QSS included in 
Table~\ref{table1} indicates that ${\cal O}_{\rm cut}$ in the range of 0.90 
to 0.95 is optimal. 
With the parameters used in this work, Eq.~(\ref{eq:AC3}) 
yields $M \sim 10^{21}$ for 
${\cal O}_{\rm cut} = 0.90$ and $M \sim 10^{22}$ for 0.95. 
If, instead of using ${\cal N}^\dag = 6$ we assume that the number of
genotypes at any time is near 8, as indicated by the top curve in
Fig.~\ref{fig:RunAve}, these
estimates are increased by six orders of magnitude. 

%\bibliography{evol.bib}

\begin{thebibliography}{10}
\bibitem{FEAS}
We constructed
  the arbitrary, feasible communities by the same method described in 
  Ref.~\protect\cite{TREG79}: 
  from a larger set of randomly selected genotypes we eliminated those
  with the most negative $n_I^*$ until all remaining $n_I^* > 0$. The resulting
  communities have ${\cal N} \in [2,5]$ with a mean of 2.5.
\bibitem{ENDNOTE}
 We define an $n$th neighbor of a community of $\cal N$ resident species as a
  genotype that can be created from one or more of the resident genotypes by
  flipping $n$ bits (Hamming distance $n$), and which is not also a member of
  the original community or a neighbor of order $<n$. If an $n$th neighbor can
  be created by flipping $n$ bits in more than one of the resident genotypes,
  it is counted with the corresponding multiplicity in the histograms shown in
  Fig.~\protect\ref{fig:eigenval}.
\bibitem{PSD}
 The PSD of a variable $y(t)$, where $t$ takes $N$ discrete values with an
  increment $\Delta$, is here defined as $Y(f) = (\Delta / N) | \sum_t [y(t) -
  \bar{y}] \exp (2 \pi i f t) |^2$, where $\bar{y}$ is the average of $y(t)$
  and $f = n/(N \Delta)$ with $n = -N/2, ..., N/2$. It was calculated with
  subroutine {\tt spctrm} from Ref.~\protect\cite{PRES92}, using a Welch window
  and variance reduction as described there.
\bibitem{NOTEb}
Our simulations with sampling every generation showed, as expected,
that the PSD does not have any simple shape for frequencies between the
Nyquist frequencies of 1/32, corresponding to sampling every
16 generations, and 1/2, corresponding to sampling every generation.
\end{thebibliography}
%\bibliographystyle{prsty}
%\bibliographystyle{unsrt}

%\clearpage

% Tables follow here. 
%%%%%%%%%%%%%%%%%%%%%%%%% TABLE1 %%%%%%%%%%%%%%%%%%%%%%%%%%%%
%\end{multicols}
%\begin{multicols}{2}
\begin{table}[ht]
\caption[]{
Composition and lifetime statistics of ten QSS that lasted
at least 20\,000 generations in a particular simulation run 
of $10^6$ generations. The QSS are listed in order of increasing mean lifetime. 
Columns one through four 
give the genotype labels and, in parentheses, the initial 
populations [the fixed-point populations
given by Eq.~(\protect\ref{eq:fix5})] for 
the $\cal N$ genotypes in each QSS. Columns five and
six give the population-weighted mean and standard deviation of the 
${\cal N}({\cal N}-1)/2$ Hamming distances between the genotypes in the initial 
community, 
Eqs.~(\protect\ref{eq:avHamm}) and~(\protect\ref{eq:devHamm}), respectively.
Columns seven and eight give the mean and standard deviations of the
${\cal N}({\cal N}-1)$ offdiagonal  
interaction matrix elements $M_{IJ}$ between the 
genotypes in the initial community, respectively. 
Columns nine and ten give the mean and standard deviations of the
lifetimes, obtained from 300 independent escapes for each QSS. 
See discussion in Secs.~\protect\ref{sec:Stab} and~\protect\ref{sec:Life}. 
}
\begin{tabular}{| r   r | r   r | r   r | r   r | c | c | c | c | c | c |}
\hline
\multicolumn{12}{| c |}{Initial QSS} & & 
\\
\multicolumn{8}{| c |}{Genotype label, $I \;$ (Population, $n_I^*$)}  
& $\langle {H} \rangle$ 
& $\sigma_{H}$ 
& $\langle M_{IJ} \rangle$ 
& $\sigma_{M_{IJ}}$ 
& $\langle {\rm Lifetime} \rangle$ 
& $\sigma_{\rm Lifetime}$
\\
\hline
5180 & (1055) & 5692 & (1506) & 7272 &  (682) &      &        & 
2.79 & 2.26 & 0.79 & 0.16 & 11\,366.2 & 11\,093.4  
\\
\hline
4251 & (1051) & 6275 & (1077) & 6283 & (1003) &   &    & 
2.02 & 1.01 & 0.70 & 0.16 & 12\,261.4 & 11\,634.4  
\\
\hline
2836 & (1624) & 2982 & (1472) &      &        &      &        & 
4.00 & 0.00 & 0.90 & 0.06 & 12\,277.7 & 19\,708.3  
\\
\hline
7135 &  (995) & 7357 &  (909) & 8191 & (1201) &      &        & 
3.77 & 1.96 & 0.67 & 0.23 & 19\,289.9 & 17\,865.6  
\\
\hline
4260 & (1212) & 4518 &  (979) & 5285 & (1078) &   &    & 
1.91 & 0.99 & 0.81 & 0.16 & 20\,216.3 & 20\,056.4  
\\
\hline
4244 & (1034) & 6164 & (1063) & 6196 &  (330) & 6676 &  (786) & 
1.99 & 0.91 & 0.67 & 0.30 & 35\,057.9 & 57\,527.9  
\\
\hline
3334 &  (915) & 3402 & (1298) & 3403 & (1094) &      &        & 
2.45 & 1.54 & 0.83 & 0.13 & 39\,577.2 & 44\,935.3  
\\
\hline
1122 & (1308) & 1146 &  (987) & 2149 & (1076) &   &    & 
4.50 & 2.43 & 0.89 & 0.07 & 39\,972.4 & 38\,328.8  
\\
\hline
7380 & (1159) & 7388 &  (683) & 7412 & (1406) &      &        & 
1.28 & 0.55 & 0.79 & 0.16 & 62\,186.5 & 67\,521.7  
\\
\hline
5860 &  (671) & 7397 & (1407) & 7653 &  (473) & 7909 &  (694) & 
1.95 & 1.17 & 0.71 & 0.25 & 80\,821.8 & 85\,716.1  
\\
\hline
 \end{tabular}
\label{table1}
\end{table}
\clearpage

%\begin{multicols}{2}
%%%%%%%%%%%%%%%%%%%%%%%%%%%%Figures yyy %%%%%%%%%%%%%%%%%%%%%%%%%%%%%%%%%%

\vspace{1.0truecm}

%\clearpage 

%\clearpage 

%\end{multicols}

\end{document}